\documentclass[twocolumn,preprintnumbers,prd,nofootinbib,superscriptaddress]{revtex4-1}

\usepackage{amsmath,amssymb,amsfonts,slashed}
\usepackage{hyperref,url,breakurl}
\usepackage{multirow}
\usepackage{graphicx}
\usepackage{mathtools,xparse}
\usepackage{placeins}

\def\bra#1{\langle #1 |}
\def\ket#1{| #1 \rangle}
\def\inner#1#2{\langle #1 | #2 \rangle}
\def\app#1#2{%
  \mathrel{%
    \setbox0=\hbox{$#1\sim$}%
    \setbox2=\hbox{%
      \rlap{\hbox{$#1\propto$}}%
      \lower1.1\ht0\box0%
    }%
    \raise0.25\ht2\box2%
  }%
}
\DeclarePairedDelimiter{\norm}{\lVert}{\rVert}
\newcommand{\I}{\textbf{\textit{I}}}
\newcommand{\X}{\textbf{\textit{X}}}
\newcommand{\Y}{\textbf{\textit{Y}}}
\newcommand{\Z}{\textbf{\textit{Z}}}

\begin{document}

\title{Classical Spacetimes as Amplified Information 
 in Holographic Quantum Theories}

\author{Yasunori Nomura}
\affiliation{Berkeley Center for Theoretical Physics, Department of Physics, 
  University of California, Berkeley, CA 94720, USA}
\affiliation{Theoretical Physics Group, Lawrence Berkeley National 
  Laboratory, Berkeley, CA 94720, USA}
\affiliation{Kavli Institute for the Physics and Mathematics of the 
 Universe (WPI), The University of Tokyo Institutes for Advanced Study, 
 Kashiwa 277-8583, Japan}
\author{Pratik Rath}
\affiliation{Berkeley Center for Theoretical Physics, Department of Physics, 
  University of California, Berkeley, CA 94720, USA}
\affiliation{Theoretical Physics Group, Lawrence Berkeley National 
  Laboratory, Berkeley, CA 94720, USA}
\author{Nico Salzetta}
\affiliation{Berkeley Center for Theoretical Physics, Department of Physics, 
  University of California, Berkeley, CA 94720, USA}
\affiliation{Theoretical Physics Group, Lawrence Berkeley National 
  Laboratory, Berkeley, CA 94720, USA}

\begin{abstract}
We argue that classical spacetimes represent amplified information in 
the holographic theory of quantum gravity.  In general, classicalization 
of a quantum system involves amplification of information at the cost 
of exponentially reducing the number of observables.  In quantum gravity, 
the geometry of spacetime must be the analogously amplified information. 
Bulk local semiclassical operators probe this information without 
disturbing it; these correspond to logical operators acting on code 
subspaces of the holographic theory.  From this viewpoint, we study 
how bulk local operators may be realized in a holographic theory of 
general spacetimes, which includes AdS/CFT as a special case, and deduce 
its consequences.  In the first half of the paper, we ask what description 
of the bulk physics is provided by a holographic state dual to a 
semiclassical spacetime.  In particular, we analyze what portion 
of the bulk can be reconstructed as spacetime in the holographic theory. 
The analysis indicates that when a spacetime contains a quasi-static 
black hole inside a holographic screen, the theory provides a description 
of physics as viewed from the exterior (though the interior information 
is not absent).  In the second half, we study how and when a semiclassical 
description emerges in the holographic theory.  We find that states 
representing semiclassical spacetimes are non-generic in the holographic 
Hilbert space.  If there are a maximal number of independent microstates, 
semiclassical operators must be given state-dependently; we elucidate 
this point using the stabilizer formalism and tensor network models. 
We also discuss possible implications of the present picture for the 
black hole interior.
\end{abstract}

\maketitle

\tableofcontents

\section{Introduction}
\label{sec:intro}

Emergence of classical spacetimes from the fundamental theory of quantum 
gravity is an important problem.  In general, classicalization of a 
quantum system involves a large reduction of possible observables. 
Suppose the final state of a scattering experiment is $c_A \ket{A} 
+ c_B \ket{B}$, where $\ket{A}$ and $\ket{B}$ are two possible particle 
states.  In principle, one can measure this state in any basis in the 
space spanned by $\ket{A}$ and $\ket{B}$.  Classicalization caused by 
the dynamics, however, makes this state evolve into a superposition 
of two classical worlds of the form $c_A \ket{AAA \cdots} + c_B \ket{BBB 
\cdots}$, in which the information about the final particles is amplified 
in each branch~\cite{q-Darwinism,q-Darwinism-2,Nomura:2011rb}.  In 
these classicalized worlds, the appropriate observable is only a binary 
question, $A$ or $B$, instead of continuous numbers associated with 
$c_A$ and $c_B$.  At the cost of this reduction of observables, however, 
the information $A$ and $B$ is now robust---it can be probed by many 
physical entities of the system, and hence is classical.  We note that 
the information amplified may depend on the state, e.g.\ the configuration 
of a detector.  (You can imagine $\ket{A}$ and $\ket{B}$ being the spin 
up and down states of a spin-$1/2$ particle.)  Given a state, however, 
the amount of information amplified is only an exponentially small 
subset of the whole microscopic information.

In quantum gravity, the information of the semiclassical spacetimes must 
be analogously amplified.  At the level of a semiclassical description, 
this information appears in the two-point functions of quantum 
field operators (a class of operators defined in code subspaces 
of the holographic theory~\cite{Almheiri:2014lwa,Pastawski:2015qua,%
Harlow:2016vwg}).  At the fundamental level, this arises mainly 
from entanglement entropies between the holographic degrees of 
freedom~\cite{Ryu:2006bv,Hubeny:2007xt,Lewkowycz:2013nqa}.  Note that 
entanglement entropies are numbers, so they comprise only an exponentially 
small fraction of the whole quantum information that the fundamental 
degrees of freedom may have, and hence the corresponding information 
may appear multiple times, e.g., in the propagators of different low 
energy fields.  This implies, in particular, that spacetime exists 
only to the extent that we can erect the corresponding code subspace 
in which the notion of local bulk operators can be defined.

In this paper, we pursue this picture in the context of a holographic 
theory for general spacetimes developed in Ref.~\cite{Nomura:2016ikr} 
(which includes AdS/CFT as a special case).  Key assumptions in our 
analyses are
\begin{itemize}
\item[(i)]
The holographic theory has degrees of freedom that appear local at 
lengthscales larger than a cutoff $l_{\rm c}$.  When a semiclassical 
description is available, the effective density of these degrees of 
freedom is $1/4$ in units of the bulk Planck length.
\item[(ii)]
If a holographic state represents a semiclassical spacetime, the area of 
the minimal area extremal surface (the HRT surface~\cite{Hubeny:2007xt}) 
anchored to the boundary of a region $\Gamma$ on a leaf $\sigma$ of a 
holographic screen gives the entanglement entropy of $\Gamma$ in the 
holographic theory~\cite{Sanches:2016sxy}.
\item[(iii)]
A quantum mechanical version of the statement (ii) above, analogous 
to those obtained/conjectured in the AdS/CFT case~\cite{Faulkner:2013ana,%
Engelhardt:2014gca}, is valid.
\end{itemize}
In Ref.~\cite{Nomura:2016ikr}, a few possible structures for the 
holographic Hilbert space have been discussed, consistent with these 
assumptions.  Our analyses in this paper, however, do not depend on the 
details of these structures, so we will be mostly agnostic about the 
structure of the holographic Hilbert space beyond (i)--(iii) above.

We emphasize that the items listed above, especially (ii) and (iii), 
are assumptions.  They are motivated by bulk reconstruction in AdS/CFT, 
but for general spacetimes their basis is weaker.  However, the 
structures in (ii) and (iii) do not seem to be particularly tied 
to the asymptotic AdS nature~\cite{Harlow:2016vwg,Hayden:2016cfa}, 
and there are analyses suggesting that they may indeed apply to 
more general spacetimes~\cite{Sanches:2016sxy,Miyaji:2015yva}.  Our 
philosophy here is to adopt them as guiding principles in exploring 
the structure of the (putative) holographic theory of general spacetimes. 
In particular, we investigate what bulk spacetime picture the general 
holographic theory provides and how it may arise from the fundamental 
microscopic structure of the theory.

Our analyses of these issues are divided into two parts.  In the first 
part, we study the question:\ given a holographic state that represents 
a semiclassical spacetime,%
\footnote{By a semiclassical spacetime, we mean a curved manifold on 
which a low energy effective field theory can be erected.  A holographic 
state representing a semiclassical spacetime, however, does not 
necessarily describe the whole spacetime region in the interior 
of the holographic screen.}
what description of the bulk physics does it provide?  For this 
purpose, we employ the tool developed by Sanches and Weinberg in 
AdS/CFT~\cite{Sanches:2017xhn}, which allows us to identify the 
region in the bulk described by a local semiclassical field theory. 
To apply it in our context, however, we need an important modification. 
To describe a general spacetime, it is essential to fix a reference 
frame, which corresponds to choosing a gauge for the holographic 
redundancy~\cite{Nomura:2011rb}.  In the bulk picture, this amounts to 
erecting a specific holographic screen with definite time slicing.  In 
fact, this time slicing has a special significance~\cite{Bousso:2015mqa}:\ 
it is the preferred time foliation in the sense that other foliations 
of the same holographic screen do not lead to equal-time hypersurfaces 
that satisfy the defining characteristic of leaves (i.e.\ marginal surfaces).

This leads us to propose that the holographic description of a general 
spacetime in a given reference frame provides a local field theoretic 
description in the region consisting of a point $p$ that can be 
written as
\begin{equation}
  p = \bigcap_\Gamma {\rm EW}(\Gamma),
\label{eq:spacetime}
\end{equation}
where ${\rm EW}(\Gamma)$ is the entanglement wedge~\cite{Wall:2012uf,%
Headrick:2014cta} of $\Gamma$, and $\Gamma$ must be chosen from spatial 
regions on leaves of the holographic screen in the given reference frame. 
We find that this criterion allows us to reconstruct most of the region 
inside the holographic screen for regular spacetimes, including some 
entanglement shadows:\ regions which the HRT surfaces do not probe. 
In AdS/CFT, the region reconstructable in this way seems to agree with 
the region obtained in Ref.~\cite{Sanches:2017xhn} using the analogous 
criterion, in which $\Gamma$ is chosen from the set of all the 
codimension-one achronal submanifolds of the AdS boundary.%
\footnote{This statement applies if the topology of the boundary space 
 is simply connected as we focus on in this paper.  If it is not, in 
 particular if the boundary space consists of disconnected components 
 as in the case of a two-sided black hole, then the two procedures 
 lead to different physical pictures.  This will be discussed 
 in Ref.~\cite{NRS}.}

We show that for a point $p$ to be reconstructable, it is sufficient that 
all the future-directed and past-directed light rays emanating from $p$ 
reach outside the entanglement shadow early enough.  We also argue that 
for $p$ to be reconstructable, at least one future-directed {\it and} 
past-directed light ray from $p$ must escape the shadow region.  This 
latter condition implies that the interior of a black hole cannot be 
reconstructed as local spacetime (except in transient periods, e.g., just 
after the formation), since the horizon of a quasi-static black hole serves 
as an extremal surface barrier~\cite{Engelhardt:2013tra}.  On the other 
hand, the analyses of Refs.~\cite{Jafferis:2015del,Dong:2016eik} 
suggest that the information about the interior is somehow contained 
in the holographic state, since the entanglement wedges of leaf regions 
cover the interior.  We interpret these to mean that the description 
of a black hole provided by the holographic theory is that of a distant 
picture:\ the information about the interior is contained in the 
stretched horizon degrees of freedom~\cite{Susskind:1993if} whose 
dynamics is not described by local field theory in the bulk.

This does not exclude the possibility that there is an effective 
description that makes a portion of the interior spacetime manifest 
by appropriately rearranging degrees of freedom.  We expect that such 
a description, if any, would be possible only at the cost of the local 
description in some other region, and it would be available only for a 
finite time measured with respect to the degrees of freedom made local 
in this manner.  We will discuss possible implications of our picture 
for the issue of the black hole interior~\cite{Almheiri:2012rt,%
Almheiri:2013hfa,Marolf:2013dba} at the end of this paper.

In the second part of our analyses, we study how and when a semiclassical 
description emerges in the holographic theory.  We first argue that when 
the holographic space of volume ${\cal A}$ is regarded as consisting 
of $N_{\cal A}$ cutoff-size cells, the number of degrees of freedom, 
$\ln k$, in each cell should be large.  This is because entanglement 
between different subregions is robust only when many degrees of freedom 
are involved.  When a semiclassical description is available, $\ln k$ 
is related to the strength of gravity in the bulk:
\begin{equation}
  \ln k = \frac{{\cal A}}{4 l_{\rm Pl}^{d-1} N_{\cal A}}
\quad
  (\gg 1),
\label{eq:large-k}
\end{equation}
where $l_{\rm Pl}$ is the Planck length in the $(d+1)$-dimensional 
bulk.  The large number of degrees of freedom in each cell implies that 
the holographic theory can encode information about the bulk in the 
configuration of these degrees of freedom, as well as in entanglement 
entropies between subregions.  Given that local semiclassical operators 
in the reconstructable region carry the entanglement entropy information, 
we might expect that the information about the other regions of spacetime 
is encoded mostly in the degrees of freedom within the cells.

Including the degrees of freedom in each cell, the holographic 
space can accommodate up to $e^{{\cal A}/4}$ independent microstates 
for the same semiclassical spacetime.  Our analysis indicates that 
a generic state in the holographic Hilbert space does not admit a 
semiclassical spacetime interpretation within the holographic screen. 
In other words, bulk gravitational spacetime emerges only as a result 
of non-genericity of states in the holographic Hilbert space.  Suppose 
there is a spacetime ${\cal M}$ that has $e^{{\cal A}/4}$ independent 
microstates.  Assumption~(ii) above then tells us that the microstates 
for such a spacetime ${\cal M}$ cannot form a Hilbert space---if it did, 
a generic superposition of these states would still represent ${\cal M}$ 
and yet have an entanglement structure that is different from what is 
implied by (ii).

At the leading order in $1/{\cal A}$, the space of microstates is 
{\it at most} the group space of $U(k)^{{\cal N}_{\cal A}}$, which 
preserves the entanglement structure between local degrees of 
freedom in the holographic theory.  This space is tiny compared with 
${\cal H}_{\cal A}$, i.e.\ the group space of $U(k^{{\cal N}_{\cal A}})$:\ 
$\norm{U(k)^{{\cal N}_{\cal A}}} \lll \norm{U(k^{{\cal N}_{\cal A}})}$. 
The actual space for the microstates, however, can be even smaller.

If the microstates comprise the elements of $U(k)^{{\cal N}_{\cal A}}$, 
then it has $e^{{\cal A}/4}$ independent microstates.  In this case, 
the semiclassical operators associated with these microstates must be 
state-dependent as argued by Papadodimas and Raju for the interior of 
a large AdS black hole~\cite{Papadodimas:2013jku,Papadodimas:2015jra}. 
This is because the code subspaces relevant for these microstates have 
nontrivial overlaps in the holographic Hilbert space.

What happens if microstates comprise (essentially) only a discrete 
$e^{{\cal A}/4}$ ``axis'' states?  In this case, different code subspaces 
can be orthogonal, so that one might think that semiclassical operators 
can be defined state-independently without any subtlety.  However, 
we argue that semiclassical operators still cannot be state-independent 
in this case.  This is because a semiclassical operator is represented 
redundantly on subregions of the holographic space as a result of amplifying 
the information about spacetime.  The necessity of state-dependence, 
therefore, is robust if any given spacetime ${\cal M}$ has $e^{{\cal A}/4}$ 
independent microstates.

The organization of this paper is as follows.  In 
Section~\ref{sec:framework}, we review our framework of the holographic 
theory of general spacetimes.  In Section~\ref{sec:classical}, we 
discuss the role of information amplification in classicalization.  
In Section~\ref{sec:spacetime}, we present the first part of our analyses. 
We study what portion of the bulk is directly reconstructable from 
a holographic state, for spacetimes without an entanglement shadow, 
with reconstructable shadows, and with non-reconstructable shadows. 
In Section~\ref{sec:non-generic}, we present the second part, in 
which we study how and when a semiclassical description emerges.  We 
discuss general features of the holographic encoding of spacetimes and 
non-genericity of semiclassical states.  In Section~\ref{sec:black-hole}, 
we conclude with remarks on possible implications of our picture for 
the black hole interior.

Throughout the paper, we adopt the unit in which the length 
$l_{\rm Pl}$---which corresponds to the bulk Planck length when 
the semiclassical picture is available---is set to unity.

\section{Framework}
\label{sec:framework}

The holographic degrees of freedom live in a holographic space, which can 
be identified as a leaf of the holographic screen~\cite{Bousso:1999cb} 
when the state admits a semiclassical interpretation.  For definiteness, 
we assume that the holographic redundancy is fixed in the observer 
centric manner~\cite{Nomura:2011dt,Nomura:2011rb}---the future-directed 
ingoing light rays emanating orthogonally from the leaf meet at a 
spacetime point (associated with the origin of a freely falling reference 
frame), unless these light rays hit a singularity before this happens.

The size (volume) of the holographic space changes as a function of 
time.  The Hilbert space relevant for the holographic degrees of freedom 
can thus be regarded as%
\footnote{It is possible that the direct sum structure arises only 
 effectively at the fundamental level.  It is also possible that the 
 Hilbert space of quantum gravity contains states that cannot be written 
 as elements of ${\cal H}_{\cal A}$.  These issues, however, do not 
 affect our arguments.}
\begin{equation}
  {\cal H} = \bigoplus_{\cal A} {\cal H}_{\cal A},
\label{eq:cal-H}
\end{equation}
where ${\cal H}_{\cal A}$ is the Hilbert space for the states of the 
degrees of freedom living in the holographic space of volume between 
${\cal A}$ and ${\cal A} + \delta {\cal A}$; namely, we have grouped 
classically continuous values of ${\cal A}$ into a discrete set by 
regarding the values between ${\cal A}$ and ${\cal A} + \delta {\cal A}$ 
as the same and labeling them by ${\cal A}$.  As in standard statistical 
mechanics, the precise way this grouping is done is not important 
(unless $\delta {\cal A}$ is taken exponentially small in ${\cal A}$, 
which is equivalent to resolving microstates and hence is not a 
meaningful choice).

The dimension of ${\cal H}_{\cal A}$ is given by
\begin{equation}
  \ln {\rm dim}\, {\cal H}_{\cal A} = \frac{{\cal A}}{4}\, 
    \biggl\{ 1 + O \biggl(\frac{1}{{\cal A}^{q>0}}\biggr) \biggr\}.
\label{eq:dim-H_A}
\end{equation}
This gives the upper bound of $e^{{\cal A}/4}$ on the number of 
independent semiclassical states having the leaf area ${\cal A}$. 
(The original covariant entropy bound of Ref.~\cite{Bousso:1999xy} 
only says that the number of independent semiclassical states 
is bounded by $e^{{\cal A}/2}$, since the number in each side 
of the leaf is separately bounded by $e^{{\cal A}/4}$.  In 
Ref.~\cite{Nomura:2016ikr}, it was argued that the actual bound 
might be stronger:\ $e^{{\cal A}/4}$ for states representing both 
sides of the leaf.  Our discussions in this paper do not depend 
on this issue.)

For the purposes of this paper, we focus on holographic spaces which 
have the topology of $\mathbb{S}^{d-1}$ with a fixed $d$, although we 
do not see a difficulty in extending this to other cases.%
\footnote{An interesting case is that the holographic space 
 consists of two $\mathbb{S}^{d-1}$ with a CFT living on each 
 of them~\cite{Maldacena:2001kr}.}
This implies that the holographic theory lives in $d$-dimensional 
(non-gravitational) spacetime, and we are considering the emergence 
of $(d+1)$-dimensional gravitational spacetime.  Following assumption~(i) 
in the introduction, we divide the holographic space of volume 
${\cal A}$ into $N_{\cal A} = {\cal A}/l_{\rm c}^{d-1}$ cutoff-size 
cells and consider that each cell can take $k = e^{l_{\rm c}^{d-1}/4}$ 
different states:
\begin{equation}
  {\cal H}_{\cal A} = {\cal H}_{\rm c}^{\otimes N_{\cal A}},
\label{eq:H_c}
\end{equation}
where ${\cal H}_{\rm c}$ is a $k$-dimensional Hilbert space associated 
with each cutoff cell.  Below, we focus on the regime
\begin{equation}
  {\cal A} \gg l_{\rm c}^{d-1},
\qquad
  \frac{l_{\rm c}^{d-1}}{4} \geq \ln 2,
\label{eq:regime}
\end{equation}
so that the setup is meaningful.

In the AdS/CFT case, $k \sim e^c$, where $c$ is the central charge of 
the CFT, which is taken to be large.  This implies that $l_{\rm c}$ is 
large in units of the bulk Planck length.  Indeed, the whole physics in 
a single AdS volume near the cutoff surface corresponds to physics of 
the $c$ degrees of freedom in a single cell of volume $l_{\rm c}^{d-1}$. 
This, however, does not mean that physics in a single AdS volume in the 
central region is confined to a description within a single boundary 
cell.  It is, in fact, delocalized over the holographic space, 
(mostly) encoded in the entanglement between the degrees of freedom 
in different cells.

\section{Classicalization and Spacetime}
\label{sec:classical}

In this section, we present a heuristic discussion on amplification of 
information and its relation to the emergence of spacetime.

As discussed in the introduction, classicalization of a quantum system 
involves amplification of information at the cost of reducing the amount 
of accessible information.  To illustrate this, consider that a detector 
interacts with a quantum system
\begin{equation}
  \ket{\Psi_{\rm s}} = c_A \ket{A} + c_B \ket{B}.
\label{eq:NR-1}
\end{equation}
The configuration of the detector can be such that it responds differently 
depending on whether the system is in $\ket{A}$ or $\ket{B}$.  The state 
of the system and detector after the interaction is then
\begin{equation}
  \ket{\Psi_{{\rm s}+{\rm d}}} 
  = c_A \ket{A} \ket{d_A} + c_B \ket{B} \ket{d_B},
\label{eq:NR-2}
\end{equation}
where $\ket{d_A}$ and $\ket{d_B}$ represent the states of the detector. 
Now suppose that an observer reads the detector.  The observer's mental 
state will then be correlated with the state of the detector:
\begin{equation}
  \ket{\Psi_{{\rm s}+{\rm d}+{\rm o}}} 
  = c_A \ket{A} \ket{d_A} \ket{o_A} + c_B \ket{B} \ket{d_B} \ket{o_B},
\label{eq:NR-3}
\end{equation}
where $\ket{o_A}$ and $\ket{o_B}$ are the observer's mental states.  The 
observer may then write the result of the experiment on a note:
\begin{equation}
  \ket{\Psi_{{\rm s}+{\rm d}+{\rm o}+{\rm n}}} 
  = c_A \ket{A} \ket{d_A} \ket{o_A} \ket{n_A} 
    + c_B \ket{B} \ket{d_B} \ket{o_B} \ket{n_B},
\label{eq:NR-4}
\end{equation}
where $\ket{n_A}$ and $\ket{n_B}$ are the states of the note after this 
is done.  We find that the information about the result is amplified in 
each term, i.e.\ it is redundantly encoded.  This implies that a physical 
entity can learn the result of the experiment by accessing any factor, 
e.g.\ $\ket{o_X}$ or $\ket{n_X}$ ($X=A,B$), without fully destroying 
the information about it in the world.  This signifies that the relevant 
information, i.e.\ $A$ or $B$, is classicalized---it can be shared by 
multiple entities in the system or accessed multiple times by a single 
physical object.

The above process of classicalization is accompanied by a reduction of 
the number of observables.  The original state of the system contains 
a qubit of information, given by two parameters $(\theta, \phi)$ spanning 
the Bloch sphere.  This manifests in the fact that depending on the 
configuration of the detector, one could have amplified the information 
in a basis other than $\{ \ket{A}, \ket{B} \}$.  Once a state is chosen, 
however, the amplification occurs only for a limited amount of information; 
in the above case, the only observable about the system in a classicalized 
world is a binary question, $A$ or $B$:
\begin{equation}
  \mbox{qubit: } (\theta, \phi) 
    \;\longrightarrow\; \mbox{bit: } A \mbox{ or } B.
\label{eq:reduction}
\end{equation}
This exponential reduction of the number of observables is the cost of 
making the information robust and is a consequence of the no-cloning 
theorem~\cite{Wootters:1982zz}.  We note that there is no issue of 
ambiguity of measurement basis in Eq.~(\ref{eq:NR-4}):\ the basis is 
determined by amplification.

Another example of classicalized states, analogous to each term in 
Eq.~(\ref{eq:NR-4}), is given by coherent states in a harmonic oscillator 
of frequency $\omega$
\begin{equation}
  \ket{\alpha} = e^{-\frac{1}{2}|\alpha|^2} 
    \sum_{n=0}^\infty \frac{\alpha^n}{\sqrt{n!}}\, \ket{n},
\label{eq:coherent}
\end{equation}
where $\alpha = |\alpha| e^{i\varphi}$ is a complex number with 
$|\alpha| \gg 1$, and $\ket{n}$ are the energy eigenstates:\ $H \ket{n} 
= (n+1/2)\omega \ket{n}$.  The information in $\alpha$ is amplified in 
the sense that it is robust under measurements, i.e.\ actions of creation 
and annihilation operators, up to corrections of order $1/|\alpha|^2$. 
For example, the action of a creation operator to $\ket{\alpha}$, 
$\ket{\tilde{\alpha}} \propto a^\dagger \ket{\alpha}$, does not affect 
the phase space trajectory of the oscillator at the leading order 
in $1/|\alpha|^2$:
\begin{equation}
  \bra{\tilde{\alpha}(t)} {\cal O}_\pm \ket{\tilde{\alpha}(t)} 
  = \bra{\alpha(t)} {\cal O}_\pm \ket{\alpha(t)} 
    \biggl\{ 1 + O\biggl( \frac{1}{|\alpha|^2} \biggr) \biggr\}.
\end{equation}
Here, $\ket{\alpha(t)} = e^{-iHt} \ket{\alpha}$ and similarly for 
$\ket{\tilde{\alpha}(t)}$, while ${\cal O}_+ = (a + a^\dagger)/2$ and 
${\cal O}_- = (a - a^\dagger)/2i$, 
giving
\begin{align}
  \bra{\alpha(t)} {\cal O}_+ \ket{\alpha(t)} 
  &= |\alpha| \cos(\omega t - \varphi),
\label{eq:O_+} \\
  \bra{\alpha(t)} {\cal O}_- \ket{\alpha(t)} 
  &= -|\alpha| \sin(\omega t - \varphi).
\label{eq:O_-}
\end{align}
Thus, the information in $|\alpha|$ and $\varphi$ can be said to be 
classicalized.  It is an exponentially small subset of the information 
that a generic microstate in the Hilbert space of the harmonic 
oscillator may carry.

The above example illustrates that the information amplification need 
not occur in real space.  It also suggests that the resulting classical 
states are generally overcomplete (for more complete discussion, see, 
e.g., Ref.~\cite{Yaffe:1981vf}).  Specifically, the space formed---not 
spanned---by $\ket{\alpha}$ is larger than that of $\ket{n}$.  Nevertheless, 
for $|\alpha| \gg 1$, the coherent states can be viewed as forming 
(approximate) basis states:\ they are nearly orthogonal
\begin{equation}
  |\inner{\alpha}{\alpha'}|^2 = e^{-|\alpha-\alpha'|^2} \lll 1,
\label{eq:ortho}
\end{equation}
and complete
\begin{equation}
  \frac{1}{\pi} \int\! d^2\alpha\, \ket{\alpha} \bra{\alpha} 
  = \hat{I},
\label{eq:complete}
\end{equation}
so that an arbitrary state $\ket{\psi}$ may be expanded as
\begin{equation}
  \ket{\psi} = \int\! d^2\alpha\, c_\alpha \ket{\alpha},
\label{eq:expansion}
\end{equation}
where $c_\alpha = \inner{\alpha}{\psi}/\pi$.  We note, again, that there 
is no basis ambiguity here because of the amplification.  Interpreted 
in terms of operators whose matrix elements between $\ket{\alpha}$ 
and $\ket{\alpha'}$ ($\alpha \neq \alpha'$) are suppressed, such 
as ${\cal O}_\pm$ giving $\bra{\alpha} {\cal O}_\pm \ket{\alpha'} 
= |\alpha' \pm \alpha^*|^2 e^{-|\alpha-\alpha'|^2} /4$, the state 
in Eq.~(\ref{eq:expansion}) appears as a superposition of different 
classical worlds.

In quantum gravity, we deal with the issue of classicalization in two 
steps.  We first deal with classicalization of the major degrees of 
freedom in the fundamental theory while leaving the rest as quantum 
degrees of freedom.  This can be done in each basis state, e.g.\ a 
single term in Eq.~(\ref{eq:NR-4}) and Eq.~(\ref{eq:expansion}).  The 
classicalized degrees of freedom correspond to background spacetime 
while the remaining ones are excitations on it (which we call matter, 
but also includes gravitons).  The resulting theory---the theory of 
quantum degrees of freedom on classical spacetime---is what we call 
semiclassical theory.  Since the way amplification occurs depends on 
the dynamics, what spacetime picture emerges may depend on the time 
evolution operator.  In this language, the reference frame dependence 
of formulating the holographic theory arises because there are multiple 
equivalent ways of describing the system using different time evolution 
operators.

Since classicalization leading to semiclassical theory is only partial, 
observables in the semiclassical theory are still quantum operators. 
The information classicalized in this process, i.e.\ background spacetime, 
appears in the two-point functions of these operators.  From the 
microscopic point of view, the semiclassical operators are defined 
by their actions in the code subspace~\cite{Almheiri:2014lwa,%
Pastawski:2015qua,Harlow:2016vwg}, and their two-point functions 
encode entanglement entropies between the fundamental holographic 
degrees of freedom~\cite{Ryu:2006bv,Hubeny:2007xt,Lewkowycz:2013nqa}. 
(This structure is visible clearly, e.g., in tensor network 
models~\cite{Swingle:2009bg,Pastawski:2015qua,Hayden:2016cfa}.) 
The information in entanglement entropies, and in more general 
entanglement structures, may be viewed as amplified; for instance, 
a maximally entangled state between two systems $A$ and $B$ is 
given by
\begin{equation}
  \ket{\Psi} \propto 
    \biggl( \prod_i e^{a_i^\dagger b_i^\dagger} \biggr)\, \ket{0},
\label{eq:max-ent}
\end{equation}
where $a_i \ket{0} = b_i \ket{0} = 0$, and gross features of entanglement 
between the two systems, including the entanglement entropy, are robust 
with respect to (a class of) measurements, i.e.\ operations of a limited 
number of creation and annihilation operators.  It is this robustness 
that allows us to take the probe approximation, and hence consider 
models adopting this approximation (e.g.\ tensor network models).

While classicalized information is amplified, it cannot be probed an 
infinite number of times (unless the system is infinitely large).  For 
example, if quantum measurements are performed to all the entities in 
Eq.~(\ref{eq:NR-4}), the information about the experimental result would 
be lost from the state.  In gravity, information about background spacetime 
can be probed by excitations in the semiclassical theory.  Their existence, 
however, necessarily affects the spacetime, so that having too many of 
them alters it completely.  It is interesting that two seemingly unrelated 
statements that probing geometry necessarily backreacts on spacetime 
and that quantum information is fragile under measurements are in fact 
related.  (A similar consideration also applies to the measurement of 
electric/magnetic fields.)

The precise way in which a semiclassical state and the code subspace 
associated with it emerge in the holographic theory is not yet 
understood.  Various aspects of this issue have been studied, e.g., 
in Refs.~\cite{Papadodimas:2013jku,Papadodimas:2015jra,Nomura:2016aww,%
Nomura:2016ikr,Jafferis:2017tiu,Qi:2017ohu}, including the dependence 
of the code subspace on a semiclassical state and the possible 
overcomplete nature of the semiclassical states.  This issue will 
be the subject of our study in Section~\ref{sec:non-generic}.

We stress that since the amplified information appears only in 
correlators of semiclassical operators, microscopic information about 
the holographic degrees of freedom is said to be measured only if it 
is probed by semiclassical operators, i.e.\ transferred to excitations 
represented by these operators.  This implies that any ``gravitational 
thermal radiation,'' e.g.\ the thermal atmosphere within the zone 
of a black hole, is not ``physical'' (does not have a semiclassical 
meaning) unless it is probed by matter degrees of freedom, e.g.\ 
detected by a physical apparatus or converted into Hawking radiation 
in the asymptotic region (outside the zone).  This is, in fact, a key 
element of a proposed solution to (the entanglement argument of) the 
firewall paradox~\cite{Nomura:2014woa,Nomura:2014voa,Nomura:2016qum} 
and the Boltzmann brain problem~\cite{Nomura:2015zda} (see 
also~\cite{Boddy:2014eba}).

\section{Reconstructing Spacetime}
\label{sec:spacetime}

In a holographic theory for general spacetimes, it is important to 
choose a reference frame to obtain a description in which the redundancy 
associated with holography (and complementarity) is fixed.  As we 
will see below, reconstructing spacetime through our method generally 
requires knowledge about the holographic state at different times. 
(For an analysis of spacetime regions reconstructed from a single 
leaf, see the appendix.)  Suppose that the state represents a 
semiclassical spacetime, at least for a sufficiently long time period. 
We are interested in knowing what portion of the spacetime is directly 
reconstructable from such a state.  In other words, we want to know 
what kind of bulk spacetime description the holographic theory provides.

For this purpose, we first define the entanglement wedge~\cite{Wall:2012uf,%
Headrick:2014cta,Sanches:2016sxy} in the form applicable to general 
spacetimes.  Let $\Gamma$ be a (not necessarily connected) region on 
a leaf, and let $E(\Gamma)$ be the HRT surface (appropriately generalized 
to include higher order effects):\ the bulk codimension-two surface 
anchored to the boundary of $\Gamma$, $\partial E(\Gamma) = \partial 
\Gamma$, extremizing the generalized entropy~\cite{Engelhardt:2014gca}.%
\footnote{We do not expect that a homology 
 constraint~\cite{Headrick:2007km,Headrick:2014cta} plays an 
 important role in our discussion, since we consider the microscopic 
 description of pure states.}
The entanglement wedge of $\Gamma$ is defined as the bulk domain of 
dependence of any achronal bulk surface $\Sigma$ whose boundary is 
the union of $\Gamma$ and $E(\Gamma)$:
\begin{equation}
  {\rm EW}(\Gamma) = D_\Sigma,
\qquad
  \partial \Sigma = \Gamma \cup E(\Gamma).
\label{eq:EW}
\end{equation}
In the AdS/CFT case, the entanglement wedge can be defined either 
associated with a spatial region $\Gamma$ or its boundary domain of 
dependence, which are equivalent if we know the conditions imposed at 
the boundary.  In general spacetimes, it is important to define the 
entanglement wedge associated with a spatial region on a leaf (a 
preferred time slice in the holographic theory), since the theory 
on the holographic screen is in general not Lorentz invariant.  In 
the AdS/CFT case, this implies that we only consider spatial regions 
$\Gamma$ on equal-time hypersurfaces in a fixed time foliation (although 
different $\Gamma$'s can be regions at different times).

We note that if we change a reference frame, the set of $\Gamma$ we 
consider changes from the bulk point of view.  In general spacetimes, 
changing the reference frame corresponds to choosing a different time 
evolution operator---in the bulk language, this ends up choosing a 
different holographic screen, and hence different leaves, from which 
$\Gamma$'s are selected.  In the AdS/CFT case, changing the reference 
frame does not affect the time evolution operator, i.e.\ CFT Hamiltonian, 
because of the high symmetry of the system---it only changes the time 
foliation to another one related by a conformal transformation.  This, 
however, does not mean that we can choose $\Gamma$ to be an arbitrary 
spacelike region.  In any fixed reference frame, $\Gamma$ should be 
restricted to spatial regions on equal-time hypersurfaces of the given 
time foliation.

Going back to the issue of reconstructing spacetime, the analyses 
of Refs.~\cite{Jafferis:2015del,Dong:2016eik}, together with our 
assumption~(iii) in the introduction, suggest that the information 
in ${\rm EW}(\Gamma)$ is in general contained in the density matrix 
of $\Gamma$ in the holographic theory.  This, however, does not mean 
that all of this information can be arranged directly in the form 
represented by local operators in the bulk effective theory.  Indeed, 
we will argue below that the portion of spacetime reconstructed in 
this way is generally smaller than the union of ${\rm EW}(\Gamma)$ 
for all $\Gamma$.  This is, in fact, consonant with the picture of 
Ref.~\cite{Susskind:1993if}.  Suppose a black hole is formed dynamically. 
The region $\cup_\Gamma {\rm EW}(\Gamma)$ then contains the region 
inside the black hole, as can be seen by considering $\Gamma$ comprising 
the entire holographic screen at a late time.  This implies that the 
information about the interior is contained in the holographic theory 
in some form, but---as we will argue---not as local excitations in 
semiclassical spacetime (while keeping locality in the entire exterior 
region).  We claim that this information corresponds to what we call 
excitations on the stretched horizon in the bulk picture.

We now assert that semiclassical spacetime as viewed from a fixed reference 
frame is composed of the set of points $p$ that can be written as
\begin{equation}
  p = \bigcap_{\Gamma \in \tilde{\cal G}} {\rm EW}(\Gamma),
\label{eq:points}
\end{equation}
where $\tilde{\cal G}$ is a subset of the collection of all the spatial 
regions on all leaves, $\tilde{\cal G} \subset {\cal G} = \{ \Gamma \}$.

There are two recent papers that used similar 
constructions~\cite{Kabat:2017mun,Sanches:2017xhn}.  In 
Ref.~\cite{Kabat:2017mun}, a local bulk operator in AdS was constructed 
in CFT using bulk HRT surfaces intersecting at that point.  This, 
however, does not allow us to construct operators in an entanglement 
shadow:\ the spacetime region which the HRT surfaces do not probe 
(see below).  Our criterion is more along the lines of the construction 
in Ref.~\cite{Sanches:2017xhn}, in which entanglement wedges associated 
with all the $(d-1)$-dimensional achronal submanifolds of the AdS 
boundary were considered to construct local operators in the AdS bulk 
(including those in an entanglement shadow).  In fact, the criterion 
of Eq.~(\ref{eq:points}) can be obtained by the logic analogous 
to that given in Ref.~\cite{Sanches:2017xhn}.  We claim, however, 
that to obtain a physical description in a fixed reference frame, the 
regions to which entanglement wedges are associated must be restricted 
to those on equal-time hypersurfaces {\it in the given time foliation}. 
In the case of AdS/CFT with simply connected boundary space, we have 
not found an example in which the region given by Eq.~(\ref{eq:points}) 
and the localizable region of Ref.~\cite{Sanches:2017xhn} differ.  In 
general spacetimes, however, one must choose the set of regions $\Gamma$ 
as described here (spatial regions on leaves).  This issue is also 
important in AdS/CFT if the boundary consists of multiple disconnected 
components~\cite{NRS}.

Below, we demonstrate that the criterion given in Eq.~(\ref{eq:points}), 
with $\Gamma$ restricted to spatial regions on leaves, allows us 
to reconstruct almost the entire spacetime, except for certain special 
regions determined by the causal structure, e.g.\ the interior of 
a black hole.  We focus our analysis to the interior of the holographic 
screen, ${\cal M} \equiv \cup_\sigma F_\sigma$, whose information is 
encoded (mostly) in entanglement between subregions in the holographic 
theory~\cite{Nomura:2016ikr}.  Here, $F_\sigma$ is the union of 
all interior achronal hypersurfaces whose only boundary is $\sigma$ 
and which does not intersect with the holographic screen except at 
$\sigma$.  The exterior of the holographic screen will be commented 
on in Section~\ref{sec:non-generic}.  Throughout, we assume that 
holographic states are pure.

\vspace{-2mm}

\subsection{Spacetime without a Shadow}
\label{subsec:no-shadow}

We first note that if a bulk point is at the intersection of $d$ HRT 
surfaces $E(\Gamma_i)$ ($i=1,\cdots,d$), then it satisfies the condition 
of Eq.~(\ref{eq:points}).  This is because for each HRT surface, we can 
include $\Gamma_i$ and its complement on the leaf, $\bar{\Gamma}_i$, in 
$\tilde{\cal G}$, so that ${\rm EW}(\Gamma_i) \cap {\rm EW}(\bar{\Gamma}_i) 
= E(\Gamma_i)$.

\begin{figure}[t]
\begin{center}
  \includegraphics[width=7cm]{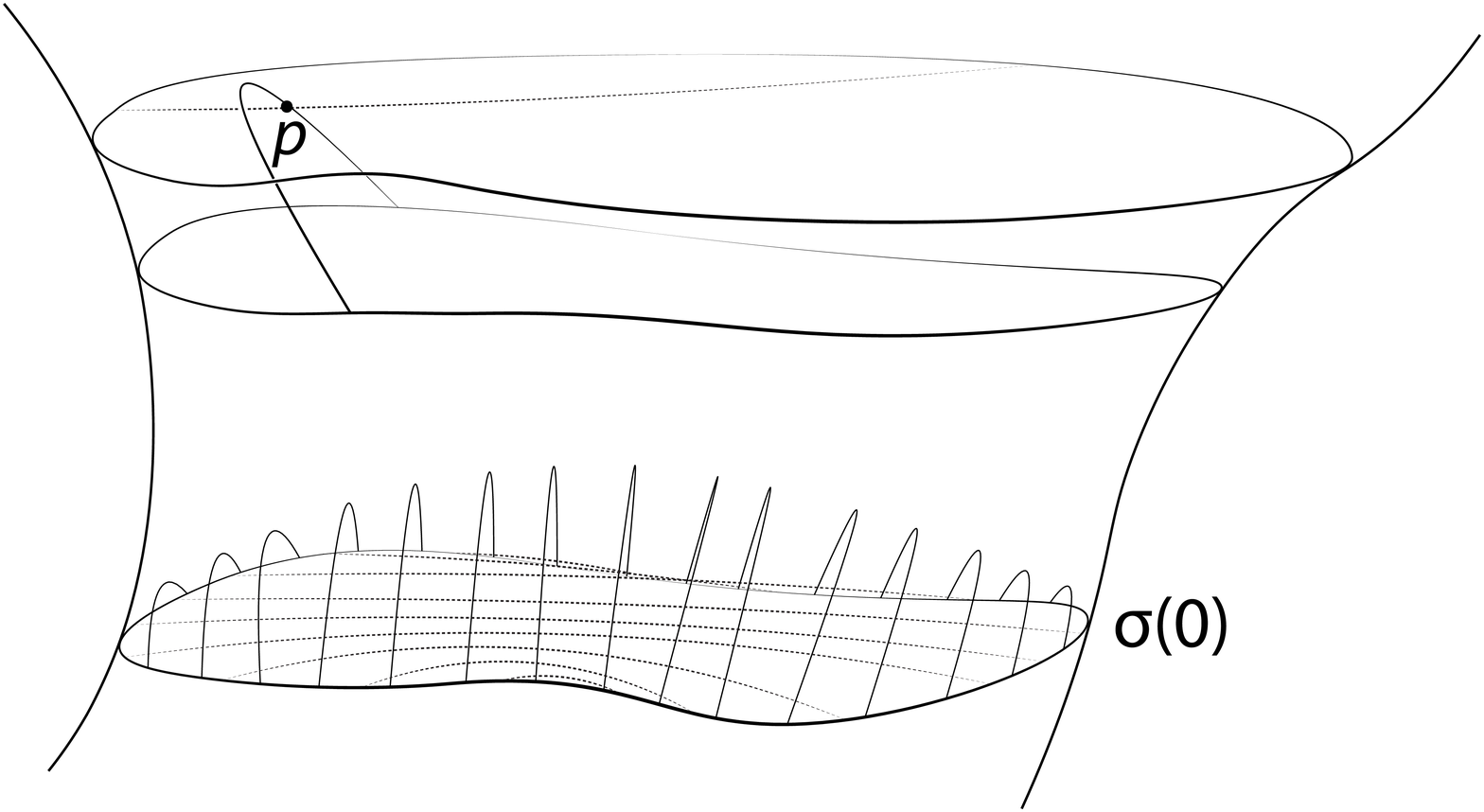}
\end{center}
\caption{If the HRT surface $E(\Gamma)$ behaves continuously under 
 a change of $\Gamma$, we can reconstruct the entire spacetime region 
 inside the holographic screen, ${\cal M}$, despite the fact that $d$ 
 families of HRT surfaces all anchored on a single leaf $\sigma(0)$ 
 do not in general span the same hypersurface.}
\label{fig:no-shadow}
\end{figure}

This implies that we can reconstruct the whole spacetime in ${\cal M}$ 
if the HRT surface $E(\Gamma)$ behaves continuously under a change 
of $\Gamma$ (i.e.\ if there is no entanglement shadow).  To show this 
explicitly, let us choose a leaf $\sigma(0)$ on the holographic screen. 
We can introduce the angular coordinates $\phi_{1,\cdots,d-1}$ on it. 
Let us now introduce the coordinates $x_j$ ($j = 1,\cdots,d$) with 
$\sum_{j=1}^d x_j^2 = 1$:
\begin{align}
  x_1     &= \cos(\phi_1),\\
  x_2     &= \sin(\phi_1) \cos(\phi_2),\\
                &\vdots \\
  x_{d-1} &= \sin(\phi_1) \cdots \sin(\phi_{d-2}) \cos(\phi_{d-1}),\\
  x_d     &= \sin(\phi_1) \cdots \sin(\phi_{d-2}) \sin(\phi_{d-1}).
\end{align}
This allows us to consider spatial regions on the leaf
\begin{equation}
  \Gamma_i^{(s)}(0) = \{ \sigma(0) \, | \, x_i \leq s \},
\label{eq:Gamma_i-s}
\end{equation}
specified by a discrete index $i = 1,\cdots,d$ and a continuous number 
$-1 \leq s \leq 1$.  Because of the continuity assumption, for each $i$ 
the corresponding HRT surfaces $E_i^{(s)}(0)$ sweep an interior achronal 
hypersurface bounded by $\sigma(0)$:
\begin{equation}
  \Sigma_i(0) \equiv \bigcup_s E_i^{(s)}(0).
\label{eq:Sigma_i}
\end{equation}
In general, the resulting $d$ hypersurfaces $\Sigma_i(0)$ ($i = 
1,\cdots,d$) are different, and the HRT surfaces contained in them 
do not intersect; see Fig.~\ref{fig:no-shadow}.

We can, however, repeat the same procedure for all different leaves 
$\sigma(\tau)$.  Here, $\tau$ is the time parameter on the holographic 
screen.  The coordinates $x_j$ on different leaves can be defined from 
those on $\sigma(0)$ by following the integral curves of a vector field 
on the holographic screen which is orthogonal to every leaf.  (Such 
a vector field was used~\cite{Sanches:2016pga} to prove that the area 
theorem of Refs.~\cite{Bousso:2015mqa,Bousso:2015qqa} is local.)

The continuity assumption then implies that for each $i$ the hypersurfaces 
$\Sigma_i(\tau)$ sweep the entire spacetime region inside the holographic 
screen, ${\cal M}$, as $\tau$ varies:%
\footnote{In the case that the holographic screen is spacelike, it seems 
 logically possible that $\Sigma_i(\tau)$ for some $i$ does not sweep 
 the entire spacetime.  We do not consider such a possibility.}
\begin{equation}
  {\cal M} = \bigcup_\tau \Sigma_i(\tau).
\label{eq:cal-M}
\end{equation}
This in turn implies that for any bulk point $p$ inside the holographic 
screen, we can find the values of $s$ and $\tau$ for each $i$, ($s_i$, 
$\tau_i$), such that the corresponding HRT surface $E_i^{(s_i)}(\tau_i)$ 
goes through $p$ (see Fig.~\ref{fig:no-shadow}).  Therefore, by choosing
\begin{equation}
  \tilde{\cal G} = \Bigl\{ \Gamma_i^{(s_i)}(\tau_i), 
    \bar{\Gamma}_i^{(s_i)}(\tau_i) \, \Bigl| \, i = 1,\cdots,d \Bigr\},
\label{eq:tilde-G_no-shadow}
\end{equation}
the point $p$ can be written as in Eq.~(\ref{eq:points}).

We note that in general, $\tau_i$ for different $i$ need not be the same. 
And yet, the region giving each entanglement wedge is on a single leaf.

\subsection{Reconstructable Shadow}
\label{subsec:reconst}

The construction described above does not apply if there is an 
entanglement shadow ${\cal S}$:\ a spacetime region which the HRT 
surfaces do not probe.  This phenomenon occurs rather generally, for 
example in spacetimes with a conical deficit~\cite{Balasubramanian:2014sra} 
or a dense star~\cite{Freivogel:2014lja}.  Here we show that a point 
$p \in {\cal S}$ may still be written as in Eq.~(\ref{eq:points}) 
if certain conditions are met.  An important point is that while 
an HRT surface $E(\Gamma)$ is always outside the shadow, the other 
part of the boundary of the entanglement wedge ${\rm EW}(\Gamma)$ 
can go into the shadow region.

\begin{figure}[t]
\begin{center}
  \includegraphics[width=8.5cm]{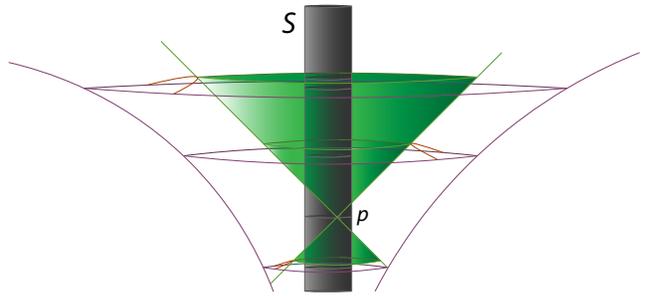}
\end{center}
\caption{A point $p$ in an entanglement shadow ${\cal S}$ can be 
 reconstructed as an intersection of entanglement wedges associated 
 with spatial regions on leaves if all the future-directed and 
 past-directed light rays emanating from $p$ reach outside the 
 entanglement shadow early enough.  Here we see that all past-directed 
 light rays escape the shadow before the first of them intersects 
 the holographic screen.}
\label{fig:shadow-1}
\end{figure}

Consider the future light cone of $p$, which we define as the subset 
of ${\cal M}$ covered by the set of future-directed light rays, 
$L^+(\Omega)$, emanating from $p$ in all directions parameterized 
by angles $\Omega = (\varphi_1, \cdots, \varphi_{d-1})$.  Similarly, 
we can consider the set of past-directed light rays $L^-(\Omega)$, 
emanating from $p$ in all directions.  Suppose all future (past) 
directed light rays escape the shadow region by the time the first 
future (past) directed light ray intersects the holographic screen 
(if at all), i.e.\ they all enter ${\cal M} \setminus {\cal S}$ 
early enough.%
\footnote{We assume that these light rays enter ${\cal M} \setminus 
 {\cal S}$ while their congruences are still expanding.  This is 
 generally true for small shadow regions.}
We now show that point $p \in {\cal S}$ can then be reconstructed 
as in Eq.~(\ref{eq:points}).  A sketch of the procedure is given 
in Fig.~\ref{fig:shadow-1}.

Let us take a point $q^+(\Omega)$ on the portion of $L^+(\Omega)$ 
in ${\cal M} \setminus {\cal S}$.  We can then find an HRT surface, 
$E^+(\Omega)$, that goes through $q^+(\Omega)$, tangent to the light 
cone there, and anchored on some leaf of the holographic screen. 
An argument is the following.  As in the previous subsection, we 
consider families of HRT surfaces anchored on $\sigma(0)$; see 
Eq.~(\ref{eq:Sigma_i}).  In the previous subsection, we considered 
$d$ such sets $E_i^{(s)}(0)$, but now we consider an infinite number 
of sets parameterized by the angular coordinates $\Phi = (\phi_1, \cdots, 
\phi_{d-1})$ on $\sigma(0)$:\ $E_{\Phi}^{(s)}(0)$.  (The corresponding 
spatial region $\Gamma_{\Phi}^{(s)}(0)$ can be taken to enlarge from 
the point specified by $\Phi$ toward its antipodal point as $s$ increases 
from $-1$ to $1$.)  Because of the entanglement shadow, these surfaces, 
$E_{\Phi}^{(s)}(0)$ ($-1 \leq s \leq 1$), do not cover the entire 
interior achronal surface bounded by $\sigma(0)$; there will be 
some hole(s).  However, by extending this to all possible leaves 
$\sigma(\tau)$, $E_{\Phi}^{(s)}(\tau)$ for each $\Phi$ will sweep 
the entire region outside the shadow:
\begin{equation}
  {\cal M} \setminus {\cal S} 
  = \bigcup_{s,\tau} E_{\Phi}^{(s)}(\tau).
\label{eq:M-S}
\end{equation}
This implies that we have a set of HRT surfaces parameterized by 
$\Phi$ that all go through $q^+(\Omega)$:
\begin{equation}
  E_{\Phi}^{(s(\Phi))}(\tau(\Phi))
  \quad \bigl(\, \ni q^+(\Omega)\, \bigr).
\label{eq:HRT-p}
\end{equation}
From these, we can choose one that is tangent to the light cone at 
$q^+(\Omega)$ because this imposes $d-1$ conditions on the $d-1$ 
parameters $\phi_1, \cdots, \phi_{d-1}$.

We therefore have the appropriate HRT surface $E^+(\Omega)$ for 
$q^+(\Omega)$, which is anchored on leaf $\sigma(\tau(\Phi))$.  There 
are two regions on this leaf that can be associated with $E^+(\Omega)$, 
which are complement with each other on the leaf.  We take the one such 
that the boundary of its entanglement wedge contains $L^+(\Omega)$, and 
we call it $\Gamma^+(\Omega)$.  We then find that the intersection of 
the entanglement wedges of $\Gamma^+(\Omega)$ for all $\Omega$ gives 
a region that is a subset of the causal future of $p$ and contains $p$:
\begin{equation}
  p \in\, \bigcap_{\Omega} {\rm EW}\bigl( \Gamma^+(\Omega) \bigr) 
  \subseteq J^+(p).
\label{eq:J+_p}
\end{equation}
Repeating the same construction for the past light cone, we obtain the 
analogous region with $+ \rightarrow -$.  Since the intersection of 
$J^+(p)$ and $J^-(p)$ is just $p$, we find that by taking
\begin{equation}
  \tilde{\cal G} = \bigl\{ \Gamma^+(\Omega),\, \Gamma^-(\Omega) 
    \, \bigl| \, \forall \Omega \bigr\},
\label{eq:tilde-G_reconst}
\end{equation}
we can write $p$ in the form of Eq.~(\ref{eq:points}).  Note that each 
region $\Gamma^+(\Omega)$ or $\Gamma^-(\Omega)$ is on a single leaf.

We conclude that to reconstruct $p$ through Eq.~(\ref{eq:points}), 
it is sufficient that all the future-directed and past-directed light 
rays emanating from $p$ reach outside the entanglement shadow early 
enough.  This condition, however, appears too strong as a necessary 
condition for the reconstruction.

In general, the bulk portion of the boundary of an entanglement wedge 
consists of three elements:\ (i) the HRT surface, (ii) null surfaces 
generated by light rays emanating orthogonally from the HRT surface, 
and (iii) caustics developed by the congruence of these light rays. 
To reconstruct a point $p$ through Eq.~(\ref{eq:points}), one of these 
elements must go through $p$.  For a point in an entanglement shadow, 
(i) is not available.  In the construction above, we have used (ii). 
We can, however, also use (iii).

\begin{figure}[t]
\begin{center}
  \includegraphics[width=7.5cm]{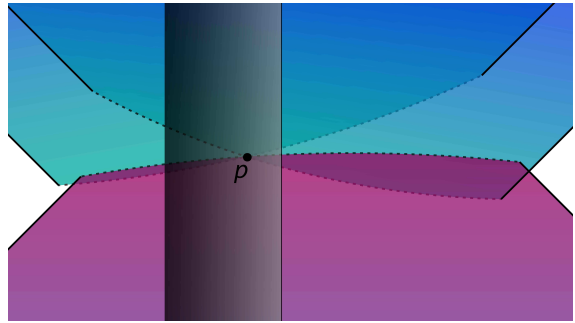}
\end{center}
\caption{A point $p$ in an entanglement shadow may be reconstructed 
 as the intersection of a finite number of entanglement wedges if 
 it is on caustics of these entanglement wedges (denoted by the 
 dotted lines).}
\label{fig:shadow-2}
\end{figure}

Consider a spatial region $\Gamma$ on a leaf.  Suppose that 
$\Gamma$ is chosen such that a caustic developed by a congruence 
of {\it past}-directed light rays emanating from $E(\Gamma)$ passes 
through $p$. Suppose also that we can find $d$ such regions, $\Gamma_i$ 
($i=1,\cdots,d$), which seems possible generically based on parameter 
counting.  Then, the intersection of ${\rm EW}(\Gamma_i)$, 
\begin{equation}
  K^+(p) = \bigcap_i {\rm EW}( \Gamma_i),
\label{eq:K+_p}
\end{equation}
forms a region which has a ``tip'' at $p$.  It therefore seems possible 
to find a region $\Gamma'$ such that a caustic developed by a congruence 
of {\it future}-directed light rays emanating from $E(\Gamma')$ passes 
through $p$, and that
\begin{equation}
  \tilde{\cal G} = \{ \Gamma_i, \Gamma' \},
\label{eq:caustics-rec}
\end{equation}
gives $p$ through Eq.~(\ref{eq:points}); see Fig.~\ref{fig:shadow-2}. 
This would allow us to reconstruct $p$ without requiring that all the 
light rays emanating from $p$ reach outside the entanglement shadow.

In all the reconstruction procedures we could consider, however, it seems 
necessary that at least one future-directed {\it and} past-directed 
light ray from $p$ escapes the entanglement shadow region.  We thus 
require this as a necessary condition for $p$ to be reconstructable. 
(This is, in fact, a very weak requirement.  In every case we considered, 
we actually needed a stronger condition.)

\subsection{Non-reconstructable Shadow}
\label{subsec:no-reconst}

The necessary condition described above has important implications. 
Suppose a black hole is formed in ${\cal M}$.  After a sufficiently 
long time, the black hole becomes quasi-static.  For any such 
black hole, HRT surfaces anchored on leaves cannot penetrate the 
horizon~\cite{Hubeny:2012ry,Engelhardt:2013tra}.  The condition described 
above then implies that the interior of a quasi-static black hole 
cannot be reconstructed directly in the holographic theory.

Two comments are in order.  First, the entanglement shadow around a 
black hole in general extends beyond the horizon (except for a large 
black hole in AdS)~\cite{Freivogel:2014lja}.  This region, however, 
is reconstructable as described in the previous subsection.  Second, 
HRT surfaces may probe the interior of the event horizon shortly 
after a black hole is formed~\cite{Hubeny:2013dea}, which allows us 
to reconstruct the region as described in Section~\ref{subsec:no-shadow}. 
After the black hole is stabilized, however, no HRT surface can 
penetrate the horizon (at least by any macroscopic distance).  The 
interior of a stabilized black hole, therefore, still cannot be 
reconstructed.

We interpret this non-reconstructability to mean that in the given 
reference frame the black hole interior is not described in terms 
of local operators in semiclassical spacetime (at least a priori; 
see below).  Suppose we (try to) represent a bulk point $p$ as in 
Eq.~(\ref{eq:points}) with some $\tilde{\cal G}$.  Let us call the 
element of $\tilde{\cal G}$ on the latest (earliest) leaf $\Gamma_+$ 
($\Gamma_-$) and the corresponding time parameter on the holographic 
screen $\tau_+$ ($\tau_-$).  Let us define $\varDelta \tau$ as the 
smallest value of $\tau_+ - \tau_-$ over the possible choices of 
$\tilde{\cal G}$:
\begin{equation}
  \varDelta \tau 
  = \underset{\tilde{\cal G}}{\rm min} \{\tau_+ - \tau_-\}.
\label{eq:delta-tau}
\end{equation}
The analyses in the previous subsections imply that for a reconstructable 
bulk point, $\varDelta \tau$ is finite.  On the other hand, for a 
non-reconstructable point, we may view that $\varDelta \tau$ is infinite 
(as the relevant light ray fails to escape the shadow region in a 
finite time).  This implies that using operators that probe (only) 
entanglement entropies between subregions, it takes an infinite time 
to resolve a point in the non-reconstructable region.  In other words, 
the effective theory for the degrees of freedom represented by these 
operators describe physics in this region as a ``vacuum degeneracy,'' 
condensed in the energy interval of $\varDelta E \sim 1/\varDelta \tau 
\rightarrow 0$.

This strongly suggests that the description obtained in the holographic 
theory is that of a distant picture for the black hole.  (Recall that 
it is the entanglement entropy structure between subregions that local 
bulk operators under consideration are mainly sensitive to.)  This, 
however, does not necessarily mean that there can be no {\it effective} 
description that makes (a portion of) the interior spacetime manifest 
by appropriately rearranging degrees of freedom.  Based on intuition 
from Ref.~\cite{Susskind:1993if}, and more recent analysis in 
Ref.~\cite{Raju:2016vsu}, we expect that such a description---if 
any---cannot keep locality in all the original spacetime region (in 
particular, outside the causal patch of a single infalling geodesic). 
In fact, we expect that any such effective description is applicable 
only for a finite time, measured with respect to the degrees of freedom 
made local in this way, reflecting the fact that the corresponding 
spacetime has a singularity.  The issue of the black hole interior 
will be discussed further in Section~\ref{sec:black-hole}.

We finally present another example of spacetime with a non-reconstructable 
region:\ an isotropic AdS cosmology.  Through a coordinate transformation, 
the interior of a future light cone $L$ in global AdS space can be written 
as an open FRW universe with the metric
\begin{equation}
  ds^2 = -dt^2 + a^2(t) \bigl( d\chi^2 + \sinh^2\!\chi\, d\Omega \bigr).
\label{eq:open-FRW}
\end{equation}
Any small perturbation makes this universe end with a big-crunch collapse 
at some time $t_*$ ($>0$), so that
\begin{equation}
  a(0) = a(t_*) = 0,
\label{eq:big-cc}
\end{equation}
where the $t=0$ hypersurface is taken to be on $L$.  In 
Ref.~\cite{Engelhardt:2013tra}, it was shown that HRT surfaces 
anchored to the AdS boundary cannot probe the region
\begin{equation}
  t > t_{\rm turn},
\label{eq:t-shadow}
\end{equation}
where $t_{\rm turn}$ is the time at which $a(t)$ becomes maximum.  Since 
any future-directed light ray emanating from a point in this region 
hits the singularity, our criterion says that this region is not 
reconstructable.  (The region $t < t_{\rm turn}$ is reconstructable 
as it is probed by HRT surfaces.)

In fact, it seems that any non-reconstructable region in realistic 
spacetimes is associated with a collapsing region (region in which 
time runs backwards in the language of Ref.~\cite{Bousso:2015mqa}) 
whose future ends in a singularity.  (We have excluded the region 
inside a past light cone in the isotropic AdS cosmology, which we 
consider ``unrealistic,'' analogous to the white hole region.)  This 
may be viewed as a quantum gravity version of cosmic censorship, 
although the surface ``hiding'' a singularity, i.e.\ that dividing 
reconstructable and non-reconstructable regions, is not necessarily 
null here.

\section{Spacetime Is Non-Generic}
\label{sec:non-generic}

We have discussed what description the holographic theory provides 
when a holographic state represents semiclassical spacetime.  Here 
we discuss how such states are embedded in the holographic Hilbert 
space.

\subsection{Holographic Encoding of Spacetime}
\label{subsec:encoding}

Consider the holographic space $\Xi_{\cal A}$ of volume ${\cal A}$, 
which consists of ${\cal N}_{\cal A}$ cutoff-size cells containing 
$\ln k$ degrees of freedom.  If a state on this space (an element 
of ${\cal H}_{\cal A}$) represents a semiclassical spacetime (or 
more precisely a snapshot of it in the holographic theory), then 
the von~Neumann entropy $S(\Gamma)$ of a subregion $\Gamma \subset 
\Xi_{\cal A}$ is related to the area of the HRT surface $E(\Gamma)$ 
as~\cite{Sanches:2016sxy}
\begin{equation}
  S(\Gamma) = \frac{1}{4} \norm{E(\Gamma)},
\label{eq:S_Sigma}
\end{equation}
ignoring the bulk matter contribution, which does not play a role in 
the discussion below.  Here, $\norm{x}$ represents the volume of the 
object $x$ (often called the area for a codimension-two surface in 
spacetime).

Suppose we take an entanglement structure $S_{\cal A} \equiv \{ S(\Gamma) 
\,|\, \forall \Gamma \subset \Xi_{\cal A} \}$ on $\Xi_{\cal A}$ implied 
by some semiclassical spacetime.  Since any unitary transformation 
acting within a single cell does not change entanglement entropies 
between subregions, this allows us to have a set of states labeled 
by the group elements of $U(k)^{{\cal N}_{\cal A}}$
\begin{equation}
  \ket{\psi^{S_{\cal A}}_z},
\quad
  z \in U(k)^{{\cal N}_{\cal A}},
\label{eq:psi-S_A}
\end{equation}
all having the same entanglement structure $S_{\cal A}$.  We expect that 
some (but not necessarily all) of these states are microstates of the 
corresponding spacetime.

In general, we expect that $k$ is large because entanglement between 
different subregions is robust (only) when many degrees of freedom are 
involved; see, e.g., Eq.~(\ref{eq:max-ent}).%
\footnote{Later, we consider perfect tensor network models, which do 
 not a priori require large bond dimensions (which one might think are 
 analogous to $k$ here).  For small bond dimensions, however, perfect 
 tensors are finely tuned:\ small perturbations would destroy their 
 absolutely maximally entangled nature.  Models with these tensors 
 can be used (only) to simulate coarse-grained structures of the 
 fundamental theory.  For large bond dimensions, this issue of stability 
 does not arise.}
In the case of AdS/CFT, i.e.\ $L_{\rm AdS}^{d-1} \ll {\cal A}$ where 
$L_{\rm AdS}$ is the AdS length, $k$ is related to the ratio of 
$L_{\rm AdS}$ to the bulk Planck length $l_{\rm Pl}$ (which we 
restore here):
\begin{equation}
  \ln k \sim \biggl( \frac{L_{\rm AdS}}{l_{\rm Pl}} \biggr)^{d-1} 
  \gg 1.
\label{eq:k-AdS}
\end{equation}
For more general spacetimes, which one may view as the case 
$L_{\rm AdS}^{d-1} \gg {\cal A}$, the meaning of $k$ is not clear, 
but one possibility is
\begin{equation}
  \ln k \sim \biggl( \frac{l_{\rm s}}{l_{\rm Pl}} \biggr)^{d-1} 
  \sim N,
\label{eq:k-general}
\end{equation}
where $l_{\rm s}$ and $N$ are the string length and the number of 
species in the low energy bulk effective theory, respectively.  We 
then expect that $k$ is also large in this case.

The degrees of freedom in $z \in U(k)^{{\cal N}_{\cal A}}$ corresponding 
to the microstates of the spacetime may contain a large amount of 
information, especially for $k \gg 1$.  Such information cannot be 
captured by entanglement between different subregions on $\Xi_{\cal A}$. 
In Section~\ref{subsec:no-reconst}, we have seen that semiclassical 
physics in a non-reconstructable region cannot be captured by entanglement 
entropies between subregions, so it is natural to conjecture that this 
physics (e.g.\ physics of the excitations of the stretched horizon) 
is encoded in these degrees of freedom.  We might suspect that physics 
outside the holographic screen may also be encoded in these degrees 
of freedom.  If this is true, the logarithmic dimension of the Hilbert 
space describing both the interior and exterior regions of the holographic 
screen is ${\cal A}/4$~\cite{Nomura:2016ikr}.  An alternative possibility 
is that the degrees of freedom describing the exterior region is 
not captured by those discussed here.  (They may not even be arranged 
locally in any space.)  In this case, the system we discuss here 
should be regarded as that responsible only for the interior region.

\subsection{Semiclassical States Are Special}
\label{subsec:special}

The number of independent states within the space of 
$U(k)^{{\cal N}_{\cal A}}$ is $k^{{\cal N}_{\cal A}} = e^{{\cal A}/4}$. 
This implies that there can be up to $e^{{\cal A}/4}$ independent 
microstates for the same semiclassical spacetime, within the uncertainties 
associated with the coarse-graining $\delta {\cal A}$, although this 
does not mean that all, or even any, semiclassical spacetimes must have 
that many independent microstates.

If a semiclassical spacetime has $e^{{\cal A}/4}$ independent microstates, 
however, it leads to the following puzzling situation.%
\footnote{One might expect that given standard de~Sitter 
 entropy~\cite{Gibbons:1977mu}, the de~Sitter FRW universe provides 
 an example of such spacetime, with $e^{{\cal A}/4}$ independent 
 microstates.  This is, however, not the case, since spacetime 
 ``disappears'' in the de~Sitter limit of the holographic FRW 
 theory~\cite{NRS}.}
Suppose there are $e^{{\cal A}/4}$ independent microstates $\ket{\psi_i}$ 
($i = 1,\cdots,e^{{\cal A}/4}$) which all correspond to (a holographic 
snapshot of) a single semiclassical spacetime ${\cal M}$ and hence have 
the same entanglement structure $S_{\cal M} = \{ S_{\cal M}(\Gamma) \,|\, 
\forall \Gamma \subset \Xi_{\cal A} \}$.%
\footnote{We expect this basis of microstates to be uncorrelated with 
 the position space basis states in holographic space and take this to 
 be the case.}
Suppose we take a superposition of $e^n$ such states
\begin{equation}
  \ket{\Psi} = \sum_{i=1}^{e^n} c_i \ket{\psi_i},
\label{eq:superposition}
\end{equation}
with comparable coefficients.  Here, $\sum_i |c_i|^2 = 1$.  If we 
compute the holographic entanglement entropy of this state in a subregion 
$\Gamma$, we generally obtain
\begin{equation}
  S(\Gamma) \neq S_{\cal M}(\Gamma).
\label{eq:S-Gamma}
\end{equation}
For small $n \ll V_\Gamma$, where
\begin{equation}
  V_\Gamma \equiv 
    {\rm min}\left\{ \norm{\Gamma}, \norm{\bar{\Gamma}} \right\},
\label{eq:V_Gamma}
\end{equation}
we find
\begin{equation}
  S(\Gamma) = S_{\cal M}(\Gamma) - \sum_{i=1}^{e^n} |c_i|^2 \ln |c_i|^2,
\label{eq:S-Gamma-2}
\end{equation}
so that the second term (classical Shannon entropy) is of order $n$, 
which is negligible compared with the first term (typically of order 
$V_\Gamma$).  If we superpose a sufficiently large number of microstates 
with $n \sim {\cal A}$, however, the entanglement structure of 
$\ket{\Psi}$, $S_\Psi \equiv \{ S(\Gamma) \,|\, \forall \Gamma 
\subset \Xi_{\cal A} \}$, can take a form unrelated with $S_{\cal M}$, 
since $\ket{\psi_i}$ form (approximately) a basis of ${\cal H}_{\cal A}$. 
Indeed, for a generic superposition with $n \approx {\cal A}/4$, 
we expect from Page's argument~\cite{Page:1993df} that $S(\Gamma) 
= V_\Gamma/4$ to a high degree.

On the other hand, one might expect that the state $\ket{\Psi}$ still 
describes semiclassical spacetime ${\cal M}$, since it is simply a 
superposition of microstates that all describe the same semiclassical 
spacetime ${\cal M}$.  If this were true, then we would find that a 
generic state describing spacetime ${\cal M}$, i.e.\ a generic state 
of the form of Eq.~(\ref{eq:superposition}), has an entanglement 
structure that has nothing to do with $S_{\cal M}$.  This would violate 
assumption~(ii) in the introduction.

We are, therefore, led to the conclusion that if a semiclassical spacetime 
has $e^{{\cal A}/4}$ independent microstates, then these states do 
not form a Hilbert space.  In fact, the space of microstates for a 
fixed semiclassical spacetime is {\it at most} the $z$ space (see 
Eq.~(\ref{eq:psi-S_A})), whose volume is tiny compared with that of 
${\cal H}_{\cal A}$:
\begin{equation}
  \norm{U(k)^{{\cal N}_{\cal A}}} \lll \norm{U(k^{{\cal N}_{\cal A}})}.
\label{eq:tiny}
\end{equation}
In fact, the actual space of microstates can be smaller. 
Note that taken at face value, the space of microstates given by 
Eq.~(\ref{eq:tiny}) is measure zero in ${\cal H}_{\cal A}$.  Our 
expressions, however, apply only at the leading order in $1/{\cal A}$, 
so we expect that the space has a nonzero ``width'' at subleading order 
in $1/{\cal A}$.  This, however, does not affect our conclusion that an 
arbitrary superposition of the form of Eq.~(\ref{eq:superposition}) 
cannot be interpreted as a semiclassical state representing ${\cal M}$.

If the microstates of a spacetime comprise the entire 
$\norm{U(k)^{{\cal N}_{\cal A}}}$ space, then what happens? 
We can use tensor network models to simulate this situation.%
\footnote{Strictly speaking, the models discussed in this and next 
 subsections apply only to the situation in which the holographic space 
 is approximately time independent, but we expect that the conclusions 
 are more general, since the time independence does not play a 
 particularly important role.}
For example, consider a tensor network obtained by contracting perfect 
tensors with some bulk legs left dangling.  By varying the perfect 
tensor chosen at each node, while keeping the network structure unaltered, 
we can generate a class of quantum error correcting codes represented 
in the holographic space.  This generates the code subspaces for all 
the geometry microstates satisfying the condition that $S(\Gamma) = 
S_{\cal M}(\Gamma)$ for all $\Gamma$.  The semiclassical (logical) 
operators obtained in this way are state-dependent, because different 
semiclassical states in different code subspaces have nontrivial overlaps 
in the holographic Hilbert space.  The same conclusion is obtained 
using random tensor networks, rather than perfect tensor networks.

A similar observation about state-dependence has been made recently 
in Ref.~\cite{Qi:2017ohu} which considered overlaps of code subspaces 
corresponding to different geometries realized in the same holographic 
space.  On the other hand, our discussion here concerns microstates 
corresponding to the same semiclassical geometry, which does not require 
different matter configurations at the semiclassical level.  The basic 
outcome for the purpose of the present discussion, however, is that 
the two situations can be treated similarly.%
\footnote{This is consonant with the picture of Refs.~\cite{Nomura:2014woa,%
 Nomura:2014voa,Nomura:2014yka} that different microstates for the 
 ``same'' spacetime (e.g.\ a single classical black hole) can/should 
 actually be viewed as states with slightly different spacetimes (black 
 holes with slightly different masses).}

The fact that the space of microstates for a fixed semiclassical 
spacetime is bounded from above by $\norm{U(k)^{{\cal N}_{\cal A}}}$ 
($\lll \norm{U(k^{{\cal N}_{\cal A}})}$) has an important implication. 
Consider a generic state in ${\cal H}_{\cal A}$.  In such a state, the 
entanglement entropy of a region $\Gamma$ is given by
\begin{equation}
  S(\Gamma) = \frac{1}{4} V_\Gamma,
\label{eq:S-typical}
\end{equation}
where $V_\Gamma$ is defined by Eq.~(\ref{eq:V_Gamma}).  This is 
because for a typical state, the reduced density matrix of a subsystem 
smaller than a half of the whole system is maximally mixed to a high 
degree~\cite{Page:1993df}.  If this state is interpreted through 
Eq.~(\ref{eq:S_Sigma}), we have
\begin{equation}
  \forall \Gamma
\quad
  \norm{E(\Gamma)} = V_\Gamma.
\label{eq:E-typical}
\end{equation}
An essentially unique way in which this happens is that the HRT surface 
anchored to $\partial \Gamma$ is $\Gamma$ itself (or $\bar{\Gamma}$, 
whichever is smaller).  The corresponding geometry in ${\cal M}$ then 
must have a horizon just inside the leaf, which serves as an extremal 
surface barrier.  Since the description of the holographic theory is 
that of the exterior picture, this state does not have any semiclassical 
spacetime inside ${\cal M}$.%
\footnote{The relation in Eq.~(\ref{eq:E-typical}) can be obtained 
 in a different way (only) if a leaf and the HRT surfaces can be mapped 
 on an extremal surface using a (infinitely) large boost transformation; 
 then the HRT surfaces lie on a null hypersurface associated with 
 the leaf.  Spacetime also disappears in this case~\cite{NRS}.  (This 
 indeed occurs in the de~Sitter limit of flat FRW universes.)}

We conclude that a generic state in the holographic Hilbert space 
does not represent a semiclassical spacetime inside the holographic 
screen.  Specifically, if the initial state of a system is generic in 
${\cal H}_{{\cal A}_0}$ and if the dynamics of the holographic theory 
is such that the state keeps being generic in ${\cal H}_{\cal A}$ 
throughout the evolution, where ${\cal A} > {\cal A}_0$, then the 
system does not admit a semiclassical spacetime interpretation 
within the holographic screen.  In this sense, we can say that bulk 
gravitational spacetime emerges only as a result of non-genericity 
of the state in the holographic Hilbert space.

We emphasize, though, that non-genericity here refers to that in the 
holographic Hilbert space without a constraint.  In particular, our 
argument does not exclude the possibility that with a specification 
of an energy range which is sufficiently lower than the cutoff, 
semiclassical states are generic among the states in that energy 
range.  This is indeed the case in standard AdS/CFT.  Similarly, 
it is possible that imposing a constraint on some other quantity 
makes semiclassical states typical within the specified class. 
Further discussion on this issue is given in Ref.~\cite{NRS}.

\subsection{State-dependence and Many Microstates}
\label{subsec:axis}

We have learned that a semiclassical spacetime inside the holographic 
screen appears only as a result of non-genericity of the holographic 
state.  We have also argued that if a semiclassical spacetime has 
$e^{{\cal A}/4}$ independent microstates, then the semiclassical operators 
are state-dependent.  This latter argument has been made by considering 
that the semiclassical microstates occupy the $U(k)^{{\cal N}_{\cal A}}$ 
space.  Here we show that the necessity of state-dependence for 
a spacetime having $e^{{\cal A}/4}$ independent microstates is even 
more robust.

The smallest possible space containing $e^{{\cal A}/4}$ independent 
microstates consists of discrete $e^{{\cal A}/4}$ ``axis'' states (with 
some small ``width'' around them).  In this case, all the different 
code subspaces can be exactly orthogonal:
\begin{equation}
  \forall a,b
\quad
  \inner{\psi^{(i)}_a}{\psi^{(j)}_b} = 0
\quad
  \mbox{for } i \neq j,
\label{eq:H_code-ortho}
\end{equation}
where $\ket{\psi^{(i)}_a}$ ($a = 1, \cdots, e^{S_{\rm code}}$) represents 
the elements of the code subspace ${\cal H}_{{\rm code},i}$ associated 
with microstate $i$.  One might then think that any semiclassical operator 
${\cal O}_X$ can be represented state-independently as
\begin{equation}
  \tilde{\cal O}_X = \bigoplus_i {\cal O}^{(i)}_X,
\label{eq:tilde-O}
\end{equation}
without any subtlety.  Here, ${\cal O}^{(i)}_X$ act ``correctly'' on 
elements of ${\cal H}_{{\rm code},i}$ but annihilate all the other states:
\begin{equation}
  \forall X, a 
\quad
  {\cal O}^{(i)}_X \ket{\psi^{(j)}_a} = 0
\quad
  \mbox{for } i \neq j.
\label{eq:O_act-ortho}
\end{equation}
Indeed, if these operators are represented in the whole holographic 
space $\Xi_{\cal A}$, then there is no obstacle in defining them as 
in Eq.~(\ref{eq:O_act-ortho}), so that we can build any semiclassical 
operator state-independently through Eq.~(\ref{eq:tilde-O}).

However, an important feature---or rather a defining property---of 
semiclassical operators is that they are represented in multiple 
different regions in the holographic space $\Xi_{\cal A}$.  Unlike 
the case of measuring a standard physical object, in quantum gravity 
there is no large external environment in which information can be 
amplified, and hence the amplification must occur ``internally'' 
within the holographic degrees of freedom given by the system.  The 
holographic theory achieves this by utilizing quantum error correction, 
amplifying information of entanglement entropies between the 
holographic degrees of freedom (which makes this information---the 
geometry---robust under operations in code subspaces).  A consequence 
of this is that operators in code subspaces, i.e.\ semiclassical 
operators, are represented in multiple subregions on $\Xi_{\cal A}$.

We now argue that the requirement of a semiclassical operator being 
represented redundantly on subregions of $\Xi_{\cal A}$ in the 
present setup prevents us from defining the operator in the form of 
Eq.~(\ref{eq:tilde-O}) acting universally on all the microstates. 
To see this, we use models given by the stabilizer formalism, which 
describes a broad class of quantum error correcting codes.  In this 
formalism, the logical states are those living in the simultaneous 
eigenspace of an abelian subgroup of the Pauli group.  For $n$ physical 
qubits, the Pauli group $G_n$ is comprised of Pauli operators which 
are a tensor product of $n$ Pauli matrices:\ $G_n = \pm \{ \I, \X, 
\Y, \Z \}^{\otimes n}$.  For qudits of higher dimensions, this can 
be appropriately generalized.

We consider that the degrees of freedom of the holographic theory are 
$n$ physical qudits.  Let ${\cal H}$ be the physical (holographic) 
Hilbert space and $T$ be an abelian subgroup of the Pauli group, which 
we consider to be {\it fixed}.  Then the states in the code subspace 
${\cal H}_{\rm code}$ can be defined as
\begin{equation}
  \ket{\psi} \in {\cal H}_{\rm code}
  \,\, \text{ iff } \,\, 
  t \ket{\psi} = \ket{\psi} \,\,\, \forall t \in T.
\label{eq:stabCodeDef}
\end{equation}
The group $T$ is called the stabilizer of the code.  We regard this code 
subspace as the Hilbert space of the semiclassical theory built on one 
of the microstates.  A class of operators that have particular significance 
are logical operators.  These operators have nontrivial action on the 
states in the code subspace and are given by elements of the Pauli group 
that commute with $T$ but are not elements of $T$.

Now, instead of Eq.~(\ref{eq:stabCodeDef}), we could have chosen any 
other of the simultaneous eigenspaces (with eigenvalues not all $+1$) 
to be our code subspace.  These eigenspaces are orthogonal and completely 
cover the full physical Hilbert space; we say that they ``tile'' 
${\cal H}$.  We identify these eigenspaces to be code subspaces 
${\cal H}_{{\rm code},i}$ associated with microstates $i = 1, \cdots, 
e^{S_{\rm micro}}$.  In this setup, each code subspace has elements 
$\ket{\psi^{(i)}_a}$ ($a = 1, \cdots, e^{S_{\rm code}}$), so that
\begin{equation}
  S_{\rm micro} \approx \frac{{\cal A}}{4},
\quad
  S_{\rm code} \ll {\cal A},
\label{eq:model-sizes}
\end{equation}
and
\begin{equation}
  S_{\rm micro} + S_{\rm code} = \ln {\rm dim} {\cal H}.
\label{eq:model-fac}
\end{equation}
It is, in fact, simple to build tensor network models realizing this 
framework.  For example, we may consider perfect tensor networks 
discussed in the previous subsection; but instead of choosing an 
arbitrary perfect tensor at each node, we now choose a tensor from 
simultaneous eigenstates of some {\it fixed} stabilizer group. 
This leads to quantum error correcting codes that have a 
particular entanglement structure $S_{\cal M}$ and correspond 
to ${\cal H}_{{\rm code},i}$ discussed above.

The quantum error correcting nature of the codes allows us to represent 
a semiclassical operator ${\cal O}_X$ for each microstate $i$ in 
various subregions $\Gamma$ in $\Xi_{\cal A}$, which we denote by 
${\cal O}^{(i)}_X(\Gamma)$.  Note that in general
\begin{equation}
  {\cal O}^{(i)}_X(\Gamma) \neq {\cal O}^{(j)}_X(\Gamma)
\quad
  \mbox{for } i \neq j,
\label{eq:O_X^i}
\end{equation}
although these operators act identically on states in their own code 
subspaces:
\begin{equation}
  \bra{\psi^{(i)}_a} {\cal O}^{(i)}_X(\Gamma) \ket{\psi^{(i)}_b} = X_{ab},
\label{eq:semicl-op}
\end{equation}
where $X_{ab}$ do not depend on $i$ or possible choices of $\Gamma$.  We 
regard that the operators obtained in this way are essentially the only 
semiclassical operators.  This realizes the situation in which the space 
of microstates consists of $e^{{\cal A}/4}$ discrete basis states (with 
some possible small ``widths'').

By construction, the operators ${\cal O}^{(i)}_X(\Gamma)$ all commute 
with the stabilizer generators
\begin{equation}
  \bigl[ t, {\cal O}^{(i)}_X(\Gamma) \bigr] = 0,
\quad
  t \in T,
\label{eq:commute}
\end{equation}
where ${\cal O}^{(i)}_X(\Gamma)$ are interpreted to be defined on the 
whole holographic space $\Xi_{\rm A}$, acting trivially on $\bar{\Gamma}$. 
This implies that actions of these operators do not send a state out of 
the code subspace it belongs to
\begin{equation}
  {\cal O}^{(i)}_X(\Gamma)\, \ket{\psi^{(j)}_a} 
  \in {\cal H}_{{\rm code},j},
\label{eq:within_i}
\end{equation}
so that the matrix elements of these operators are nonzero only between 
states in the same code subspace
\begin{equation}
  \bra{\psi^{(j)}_a} {\cal O}^{(i_1)}_{X_1}(\Gamma) \cdots 
    {\cal O}^{(i_m)}_{X_m}(\Gamma) \ket{\psi^{(k)}_b} \propto \delta_{jk}.
\label{eq:delta_jk}
\end{equation}
We also find that the matrix elements involving states and operators of 
different code subspaces have $O(1)$ entries but only with the probability 
of $e^{-S_{\rm code}}$:
\begin{align}
  & \bra{\psi^{(j)}_a} {\cal O}^{(i_1)}_{X_1}(\Gamma) \cdots 
    {\cal O}^{(i_m)}_{X_m}(\Gamma) \ket{\psi^{(j)}_b} 
  \sim O(1)
\nonumber\\
  & \qquad\qquad\qquad\qquad \mbox{with } P \sim e^{-S_{\rm code}},
\label{eq:exp^-S_code}
\end{align}
where we have normalized operators such that nonvanishing $X_{ab}$ in 
Eq.~(\ref{eq:semicl-op}) are $O(1)$.

The property of Eq.~(\ref{eq:exp^-S_code}) follows because the quantum 
error correcting code corresponding to each microstate $i$ can be viewed 
as a single large tensor having logical qudits and physical qudits as 
its indices.  The set of tensors corresponding to all the microstates 
can then be viewed as the simultaneous eigenstates of some fixed 
(stabilizer) generators acting on all these indices.  This structure 
guarantees that Eq.~(\ref{eq:exp^-S_code}) is satisfied.%
\footnote{Instead of adopting the exact stabilizer formalism as we did 
 here, we could use random tensor network models to simulate the setup 
 in which the microstates comprise $e^{{\cal A}/4}$ axis states.  To 
 do so, we can choose generic $e^{{\cal A}/4}$ codes from those obtained 
 by randomly varying the tensor at each node of a fixed network; 
 these $e^{{\cal A}/4}$ codes then approximately tile ${\cal H}$. 
 In these models, essentially all the elements in the left-hand side 
 of Eq.~(\ref{eq:exp^-S_code}) are nonzero, but they are uniformly 
 suppressed as $e^{-S_{\rm code}/2}$.  This does not change the 
 conclusion of our analysis here.}

We now see how the properties in 
Eqs.~(\ref{eq:within_i}~--~\ref{eq:exp^-S_code}) prevents us from 
having state-independent semiclassical operators.  In order for 
exactly state-independent semiclassical operators to be defined, 
${\cal O}^{(i)}_X(\Gamma)$ must satisfy
\begin{align}
\qquad
  & \bra{\psi^{(j)}_a} {\cal O}^{(i_1)}_{X_1}(\Gamma) \cdots 
    {\cal O}^{(i_m)}_{X_m}(\Gamma) \ket{\psi^{(j)}_b} 
  \neq 0
\nonumber\\
  & \qquad\qquad\qquad \mbox{only for } i_1 = \cdots = i_m = j,
\label{eq:univ-cond}
\end{align}
at least for small values of $m$.  In this case, the operators 
represented in subregion $\Gamma$
\begin{equation}
  \tilde{O}_X(\Gamma) 
  = \sum_{i=1}^{e^{S_{\rm micro}}} {\cal O}^{(i)}_X(\Gamma),
\label{eq:univ-Gamma}
\end{equation}
would become the direct sum form of Eq.~(\ref{eq:tilde-O}) and act 
correctly on all possible states of the form
\begin{equation}
  \ket{\Psi_a} = \sum_{i=1}^{e^n} c_i \ket{\psi^{(i)}_a}.
\label{eq:semicl-state}
\end{equation}
However, for $\Gamma \neq \Xi_{\cal A}$, the conditions in 
Eq.~(\ref{eq:univ-cond}) are not satisfied for all operators; 
see Eq.~(\ref{eq:exp^-S_code}).

In fact, nonzero entries in Eq.~(\ref{eq:exp^-S_code}) do not allow 
for even approximately state-independent operators.  To see this, 
consider the matrix elements
\begin{align}
  & \bra{\psi^{(j)}_a} \tilde{O}_X(\Gamma) \ket{\psi^{(j)}_b} 
  = \sum_{i=1}^{e^{S_{\rm micro}}} 
    \bra{\psi^{(j)}_a} {\cal O}^{(i)}_X(\Gamma) \ket{\psi^{(j)}_b} 
\nonumber\\
  &\qquad
  = X_{ab} + \sum_{i=1; i \neq j}^{e^{S_{\rm micro}}} 
    \bra{\psi^{(j)}_a} {\cal O}^{(i)}_X(\Gamma) \ket{\psi^{(j)}_b},
\label{eq:mat-el}
\end{align}
where we have used Eq.~(\ref{eq:semicl-op}) and $\sum_i |c_i|^2 = 1$. 
The first term is what we want, but the second term gives a much 
larger contribution $\sqrt{e^{S_{\rm micro}} e^{-S_{\rm code}}} \gg 
|X_{ab}| \sim O(1)$, where the square root in the leftmost expression 
arises because of random phases.  This implies that we cannot define 
state-independent semiclassical operators even approximately.

Note that the origin of the state-dependence is not the overlap 
between different code subspaces.  In fact, different code subspaces 
are orthogonal, Eq.~(\ref{eq:H_code-ortho}), in the present (extreme) 
setup of discrete microstates.  Semiclassical operators, however, still 
must be defined state-dependently because of the requirement of being 
represented redundantly in the holographic space.  We find that the 
necessity of state-dependence is robust in the holographic theory if 
there are $e^{{\cal A}/4}$ independent microstates for a semiclassical 
spacetime.

We thus conclude either that semiclassical states are special or that 
bulk operators are state-dependent, in which case semiclassical states 
can be generic.

\section{Black Hole Interior}
\label{sec:black-hole}

We finally discuss the issue of the black hole interior within our 
framework.  A simple description of the interior would arise if 
a portion of the holographic screen enters inside the black hole 
horizon.  However, we find this is unlikely to occur in a realistic 
setup in which the second law of thermodynamics, $d{\cal A}/d\tau > 0$, 
is obeyed.  First, the holographic screen cannot approach close 
to the singularity, since then the area of the leaf would decrease 
in time, contradicting the assumption of $d{\cal A}/d\tau > 0$. 
This leaves the possibility that a portion of a leaf enters the black 
hole and then exits or terminates.  Even in this case, however, we would 
still encounter the strange situation where the portion of a leaf inside 
the black hole has a larger area than the corresponding part of the black 
hole horizon.  We therefore assume that the holographic screen does not 
enter inside the black hole horizon (except possibly in transient periods), 
though a general proof is lacking.

This only leaves the possibility that the black hole interior can be 
described effectively by rearranging the degrees of freedom of the 
theory (which include the stretched horizon degrees of freedom identified 
in Section~\ref{sec:spacetime}).  Such a description would make 
approximate locality in a portion of the interior manifest at the 
cost of the local description in some other region (the complementarity 
picture~\cite{Susskind:1993if}).%
\footnote{In AdS/CFT, this might be done along the lines of 
 Refs.~\cite{Papadodimas:2013jku,Papadodimas:2015jra}.  We suspect 
 that the requirement of the same interior region being represented 
 redundantly, associated with amplification, might address the 
 question of why the specific set of operators considered in 
 Refs.~\cite{Papadodimas:2013jku,Papadodimas:2015jra} has a special 
 physical significance.}
We note that this rearrangement of the degrees of freedom would have 
a different nature than just changing the reference frame, e.g., by 
boosting the origin of a freely falling reference frame with respect 
to which the holographic screen is erected.  In fact, we expect that 
any effective description of the black hole interior is applicable only 
for a finite time (measured with respect to the degrees of freedom made 
local) reflecting the existence of the singularity, while the reference 
frame change would give another description of the system which does 
not have such a restriction.

What about the arguments of Refs.~\cite{Almheiri:2012rt,Almheiri:2013hfa,%
Marolf:2013dba} then, which seem to exclude even the possibility of this 
kind of (effective) description?  These arguments can essentially be 
summarized into two classes:
\begin{itemize}
\item[] {\it Entanglement argument.}
Consider an outgoing mode localized in the zone, corresponding to a 
Hawking quanta just emitted from the stretched horizon.  Unitarity 
requires this mode to be entangled with a mode representing Hawking 
radiation emitted earlier, while the smoothness of the horizon requires 
it to be entangled with the pair mode inside the horizon.  These two 
cannot both be true because of monogamy of entanglement.
\item[] {\it Typicality argument.}
Suppose we calculate the average of the number operator $\hat{a}^\dagger 
\hat{a}$ in the dual field theory over states having energies in a 
chosen range, with $\hat{a}$ corresponding to an infalling mode in the 
bulk.  The resulting number is at least of order unity, because one 
can choose a basis for these states such that they are all eigenstates 
of the number operator $\hat{b}^\dagger \hat{b}$ with $\hat{b}$ 
corresponding to a mode localized in the zone (and because the 
expectation value of $\hat{a}^\dagger \hat{a}$ in any eigenstate 
of $\hat{b}^\dagger \hat{b}$ is at least of order unity).  This implies 
that the expectation value of $\hat{a}^\dagger \hat{a}$ is of order unity 
or larger, giving firewalls, in a typical state in this energy range.
\end{itemize}

The former, entanglement argument was addressed in 
Refs.~\cite{Nomura:2014woa,Nomura:2014voa}.  At the level of 
a semiclassical description, the Bekenstein-Hawking entropy, 
$S_{\rm BH} = {\cal A}/4$, can be interpreted as the logarithm of 
the number of independent black hole states of masses between $M$ 
and $M + \varDelta M$, where $\varDelta M$ can be taken naturally as 
the inverse of the Hawking emission timescale.  Interpreted in terms 
of semiclassical operators, this information is distributed according 
to the thermodynamic entropy associated with the blue-shifted Hawking 
temperature.  This implies that while most of the information is 
concentrated near the stretched horizon, it has some spread over the 
zone.  In particular, an $O(1)$ amount of information---which is an 
$O(1/{\cal A})$ fraction of the full Bekenstein-Hawking entropy---is 
at the edge of the zone.

From the semiclassical viewpoint, Hawking emission is a process in 
which the black hole information (and energy), stored in spacetime, 
is converted into that of semiclassical excitations {\it at the edge 
of the zone} (more precisely, the region around the edge of the zone 
with the radial width of order the wavelength of emitted Hawking 
quanta).  Note that in the semiclassical viewpoint it is natural 
that the process occurs in this particular region; it is where the 
two static geometries---the near horizon, Rindler-like space and 
asymptotic, Minkowski-like space---are ``patched'' to obtain the 
full geometry.  This implies that it is incorrect to view that Hawking 
emission (and the associated information transfer) occurs through 
outgoing semiclassical excitations in the zone as envisioned in 
Refs.~\cite{Almheiri:2012rt,Almheiri:2013hfa,Marolf:2013dba}.  In 
fact, the transfer of energy and information must be viewed as occurring 
through the flux of negative energy and negative entropy, defined 
with respect to the static, Hartle-Hawking vacuum.

The typicality argument does not apply when semiclassical operators are 
given state-dependently~\cite{Papadodimas:2013jku,Papadodimas:2015jra}. 
Moreover, if the black hole microstates comprise only a subset of 
the space spanned by the independent microstates, as contemplated in 
Section~\ref{subsec:axis}, then the argument may become irrelevant 
because the black hole microstates would indeed be non-generic. 
If this is the case, then smooth black hole states would have to 
be selected dynamically.

Finally, the fact that the holographic screen does not enter the black 
hole may allow us to take the attitude that the black hole interior 
need not be described, since a measurement performed in the interior 
cannot be communicated to an external observer described directly in 
the holographic theory.  Of course, what ``happened inside'' is encoded 
indirectly in the final Hawking radiation (and in the configuration 
of the stretched horizon degrees of freedom at intermediate stages), 
which can be described appropriately in the holographic theory.

\acknowledgments

We thank Chris Akers, Venkatesa Chandrasekaran, Adrian Franco Rubio, 
Aitor Lewkowycz, Fabio Sanches, Arvin Shahbazi Moghaddam, and Sean 
Weinberg for discussions.  This work was supported in part by the National 
Science Foundation under grants PHY-1521446, by MEXT KAKENHI Grant Number 
15H05895, and by the Department of Energy (DOE), Office of Science, 
Office of High Energy Physics, under contract No.\ DE-AC02-05CH11231.

\appendix

\section{Reconstruction from a Single Leaf}
\label{app:single-leaf}

In this appendix, we study what portion of the bulk can be reconstructed 
from a {\it single} leaf, i.e.\ by Eq.~(\ref{eq:points}) with all 
$\Gamma$ on a single leaf.  We refer to three types of spacetime regions 
discussed in Sections~\ref{subsec:no-shadow}, \ref{subsec:reconst}, and 
\ref{subsec:no-reconst} as non-shadow, reconstructable shadow, and 
non-reconstructable shadow regions, respectively. 
\begin{table*}
\begin{center}
\begin{tabular}{c|cc}
  & Single leaf & Multiple leaves \\ \hline
  \multirow{2}{*}{Non-shadow} & 
    codimension-$0$ & \multirow{2}{*}{codimension-$0$} \\
    & (codimension-$1$ for $t \leftrightarrow -t$) & \\
  \multirow{2}{*}{Reconstructable shadow} & 
    \multirow{2}{*}{none} & \multirow{2}{*}{codimension-$0$} \\
  & & \\
  \multirow{2}{*}{Non-reconstructable shadow} & 
    \multirow{2}{*}{none} & \multirow{2}{*}{none} \\
    & & \\
\end{tabular}
\end{center}
\caption{The dimensions of bulk spacetime regions directly reconstructable 
 from a single leaf (the left column) and multiple leaves (the right 
 column).  The entry ``none'' means that no region can be reconstructed.}
\label{tab:single}
\end{table*}

Let us first consider a spacetime point $p$ in a non-shadow region. 
We show that generically a codimension-$0$ neighborhood of $p$ can be 
directly reconstructed from a single leaf $\sigma$.  For this purpose, 
we consider a timelike geodesic, $p(\tau)$, that goes through $p$ 
at $\tau = 0$ and stays in the non-shadow region.  We then introduce 
a map $f$ from the coordinates, $\Phi = (\phi_1, \cdots, \phi_{d-1})$, 
parameterizing $\sigma$ to $\tau$ as follows.  For each $\Phi$, we 
consider a continuous family of HRT surfaces associated with subregions 
of $\sigma$ which are taken to enlarge naturally from the point specified 
by $\Phi$ toward its antipodal point.  Generically, one of these HRT 
surfaces intersects with $p(\tau)$ at some $\tau$, which we take as 
the image of the map:\ $\tau = f(\Phi)$.  Now, we can choose $\sigma$ 
so that ${\rm max}_\Phi f(\Phi) > 0$ and ${\rm min}_\Phi f(\Phi) < 0$. 
The set satisfying $f(\Phi) = 0$ is generally codimension-$1$ in the 
space of $\Phi$, hence there are many such $\Phi$ which have an HRT 
surface passing through $p$.  This tells us that $p$ can be reconstructed 
from this single leaf.  In addition, continuity suggests that some 
interval of $p(\tau)$ is reconstructed from this leaf.  Applying this 
argument for points outside of entanglement shadows tells us that a 
single leaf generally allows for the reconstruction of a codimension-$0$ 
region of the bulk.

Note that there is a case in which the above argument does not apply. 
Suppose the spacetime is time reflection symmetric with respect 
to $\sigma$, as in the case of a static system.  Then all HRT surfaces 
anchored to $\sigma$ live on the same bulk hypersurface corresponding to 
the reflection symmetric point.  This therefore allows for reconstructing 
only a codimension-$1$ region.  The situation for reconstructing 
non-shadow regions from a single leaf is summarized in the upper-left 
entry of Table~\ref{tab:single}.

We now prove that a single leaf cannot reconstruct any portion of an 
entanglement shadow.  We will do so by contradiction.  Suppose a point 
$p$ resides within an entanglement shadow but can be reconstructed by 
a single leaf, $\sigma$.  Then there is a subregion of $\sigma$, $A$, 
such that $p$ lies on the boundary of the entanglement wedge of $A$. 
Now, suppose some subregion of $\sigma$, $B$, has nonzero overlap with 
$A$, i.e.\ $A \cap B$ is codimension-$0$ on the leaf.  Then ${\rm EW}(A) 
\cap {\rm EW}(B)$ is codimension-$0$ in the bulk, and will not be 
sufficient to localize $p$.  Therefore, there must be some subregion 
of $\sigma$, $C$, where $C \subseteq \bar{A}$ and the boundary of 
${\rm EW}(C)$ intersects $p$.  However, from entanglement wedge 
nesting, ${\rm EW}(C) \subseteq {\rm EW}(\bar{A})$, and thus it 
cannot be the case.  Note that this is a result of subregion duality, 
which tells us that the only intersection of ${\rm EW}(A)$ and 
${\rm EW}(\bar{A})$ is the HRT surface.  This implies that multiple 
leaves are necessary to reconstruct bulk regions in entanglement 
shadows.

Table~\ref{tab:single} gives a summary of directly reconstructable 
bulk regions from boundary states.

\end{document}